\documentclass[11pt]{article}
\usepackage[a4paper,margin=2.5cm]{geometry}
\usepackage{amsmath,amssymb,bm}
\usepackage{graphicx}
\usepackage{array}
\usepackage{placeins}
\usepackage{xcolor}
\usepackage{authblk}
\usepackage[numbers,sort&compress]{natbib}
\usepackage[colorlinks=true,linkcolor=blue,citecolor=blue,urlcolor=blue]{hyperref}
\usepackage[labelfont=bf,labelsep=period]{caption}
\usepackage{setspace}
\graphicspath{{figures/}}
\renewcommand{\figurename}{Fig.}

\newcommand{\U}{\bm{U}}
\newcommand{\pp}[1]{\varphi_{#1}}                              
\newcommand{\pn}[1]{\varphi^{\mathrm{norm}}_{#1}}              
\newcommand{\pnl}[2]{\varphi^{\mathrm{norm},\,\mathrm{#1}}_{#2}}
\newcommand{\Dphi}{\Delta\varphi}
\newcommand{\fr}[2]{\psi^{(#1)}_{#2}}                          
\newcommand{\Gam}[2]{\Gamma^{(#1)}_{#2}}                       
\newcommand{\ee}{\mathrm{e}}
\newcommand{\ii}{\mathrm{i}}
\newcommand{\FB}{F_{\mathrm{B}}}
\newcommand{\Fphi}{F_{\varphi}}
\newcommand{\Iv}{\bm{I}}
\newcommand{\openone}{\mathbb{I}}

\begin{document}
\title{Unveiling gauge-invariant on-chip phases from light intensity}

\author[1,*]{Huiwen Li}
\author[1]{Andrea Cataldo}
\author[1]{Rohan Yadgirkar}
\author[2,3,4]{Jun Gao}
\author[1,*]{Ali W. Elshaari}
\affil[1]{Department of Applied Physics, KTH Royal Institute of Technology, Albanova University Centre, Roslagstullsbacken 21, 106 91 Stockholm, Sweden}
\affil[2]{School of Optical and Electronic Information, Huazhong University of Science and Technology, Wuhan 430074, China}
\affil[3]{Research Center for Cross-Scale Global-Scale Multimodal Optical Communication Discipline and Technology, Wuhan 430074, China}
\affil[4]{Optics Valley Laboratory, 1037 Luoyu Road, Wuhan 430074, People's Republic of China}
\affil[*]{huiwenl@kth.se; elshaari@kth.se}
\date{}

\maketitle

\begin{abstract}
Physics-informed digital twins fitted to output powers increasingly program photonic processors, but these powers do not fix the implemented phase, which determines how light from different inputs interferes. Here we show that a twin trained on intensities alone predicts the gauge-invariant phase of programmed transformations, and that this prediction can be used to program the phase. On a $12\times12$ mesh, with phases measured by interference protocols formed by the mesh itself, intensity-only and phase-aware programs from random starts reach the same intensity fidelity but phase fidelities of $0.690$ and $0.996$, both as predicted by the twin. The prediction holds for blocks up to $10\times10$. With coherent inputs, the programmed phase determines the outputs: phase-aware programming reduces the direction-of-arrival error of an on-chip Butler matrix from $29^\circ$ to $2.0^\circ$. Power measurements alone thus yield a phase-calibrated model of the mesh, without phase data in training or optics beyond the chip.
\end{abstract}

Programmable photonic processors route light through meshes of Mach--Zehnder interferometers (MZIs) whose phases are set by heaters~\cite{Reck1994,Clements2016,Bogaerts2020}. A universal mesh of $N$ modes can implement any $N\times N$ unitary transformation, and such meshes are used in optical neural networks~\cite{Shen2017,Pai2023} and signal processing~\cite{Bogaerts2020}. Fabrication errors shift the phases and coupling ratios of a fabricated chip from their nominal values, so a setting computed for an ideal mesh does not give the target transformation, and the chip must be calibrated~\cite{Bandyopadhyay2021,Hamerly2022}. A common approach fits a physical model of the chip, often called a digital twin, to measured output powers and programs the chip through this model~\cite{Fyrillas2024,Kondratyev2026,Cataldo2026PRL}. Trained on powers alone, such models reproduce the measured intensities with high fidelity; on a 12-mode mesh, an intensity-trained model programs Haar-random unitaries with an average amplitude fidelity of 99.77\,\%~\cite{Fyrillas2024}. Intensities, however, do not fix the phase of the implemented transformation. The complex Hadamard matrices of order four, for example, form a continuous family whose elements all have power $1/4$~\cite{TadejZyczkowski2006}, so an intensity-only objective cannot tell them apart. The phase determines how light from different inputs interferes, and it matters whenever light enters several inputs coherently: in complex-valued optical neural networks, whose network weights carry phases as well as amplitudes and whose inputs are encoded in all ports~\cite{Zhang2021CVNN,Shen2017,Pai2023}, or in beamforming networks such as the Butler matrix, which separate the directions of incoming waves by interference~\cite{ButlerLowe1961}.

Testing the phase of a programmed transformation is itself difficult. Light picks up unknown phases on its way to and from the mesh, so only combinations of matrix elements in which these port phases cancel can be measured. Such combinations have been obtained from single- and two-photon measurements~\cite{Laing2012,Dhand2016}, from coherent beams prepared outside the chip~\cite{RahimiKeshari2013,Hoch2023} and from on-chip homodyne interferometers that add a dedicated reference arm~\cite{Tang2026}, and models without physical constraints are reconstructed from intensity data with controlled input phases or from phase-resolved tomography~\cite{Bantysh2024,Kuzmin2025}. On the processor used here, the relative phases of an output state have also been read by interfering pairs of outputs in an MZI of the mesh behind the programmed block, with the phase reference taken from a calibration of each phase shifter~\cite{Xu2026Sync}. A physical model can instead be fitted to single-input powers alone, and for a non-universal interferometer with three phase shifters its predicted transfer matrices agreed with matrices measured by two-beam interference~\cite{Kondratyev2026}. Whether this holds for a universal mesh with over a hundred phase shifters, programmed in blocks of different sizes and positions, and whether the predicted phase can then be used to program the chip, has not been tested (Supplementary Note~\ref{app:related}). Here we show that the twin of a $12\times12$ Clements mesh, trained on single-input powers alone, predicts the gauge-invariant phase of $m\times m$ blocks programmed on the chip, and that minimising the difference between this prediction and the target phase in the inverse design makes the chip implement the target phase. We measure the phase with two dual-input protocols built from the MZIs in front of the block, without optics beyond the chip. The prediction holds for blocks from $3\times3$ to $10\times10$ at several positions. Finally, we prepare coherent inputs on the chip and show that the programmed phase determines the outputs, including those of an on-chip Butler matrix used for direction-of-arrival estimation.


\section*{Results}

\subsection*{Intensity-trained digital twin}

\begin{figure}[htbp]
\centering
\includegraphics[width=\textwidth]{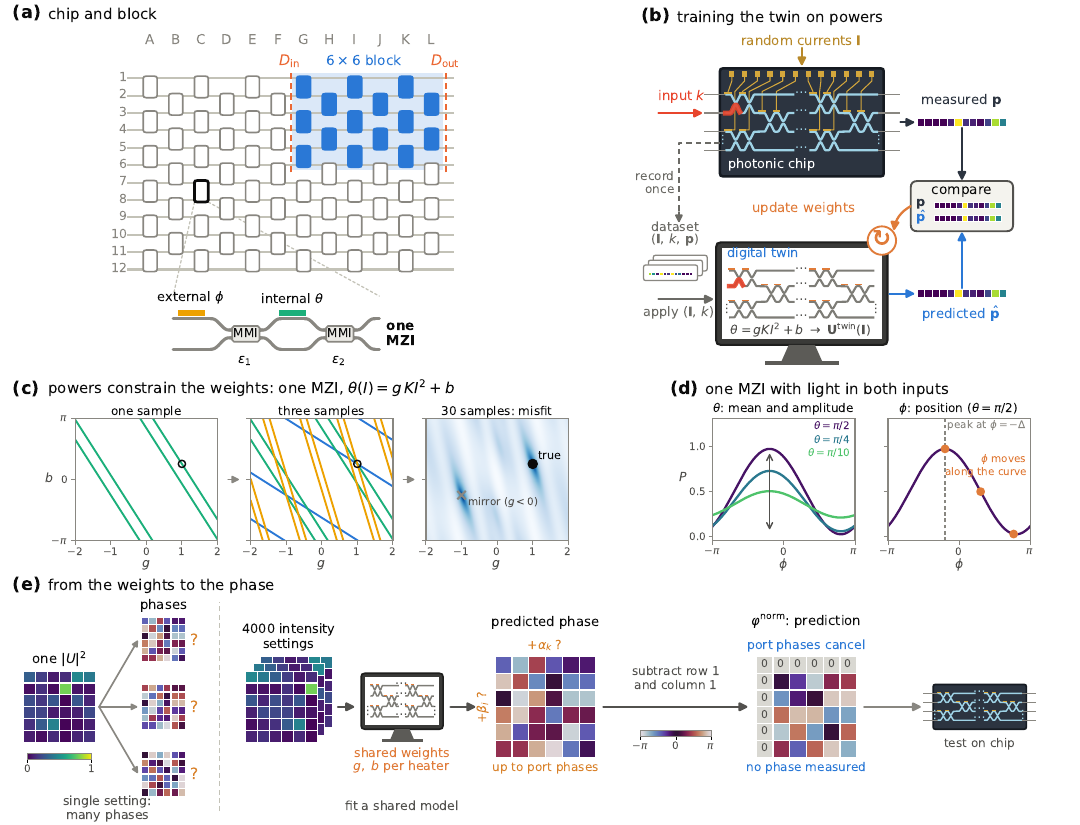}
\caption{Chip and digital twin. (a) $12\times12$ Clements mesh with layers A--L and a $6\times6$ block on ports 1--6 (blue); $\bm{D}_{\rm in}$ and $\bm{D}_{\rm out}$ are the unknown port phases before and after the block. Inset: one MZI, with the external phase $\phi$, the internal phase $\theta$ and two MMI couplers with imbalances $\epsilon_1$ and $\epsilon_2$. (b) Training: for random currents $\Iv$ and light in input $k$, the measured powers $\bm p$ are recorded once; the twin receives the same $(\Iv,k)$, predicts the powers $\hat{\bm p}$, and its weights are updated until $\hat{\bm p}$ matches $\bm p$. (c) Powers constrain the weights of one ideal MZI with $\theta(I)=gKI^2+b$: one power sample allows a family of lines in the $(g,b)$ plane, three samples cross only near a few points, and the misfit of 30 samples is lowest at the true point and at its mirror $(-g,-b)$, which $g>0$ excludes (noiseless example). (d) Bar-output power of one MZI with light in both inputs against the external phase $\phi$ (Supplementary Eq.~\eqref{eq:mzi_two_inputs}; unequal input powers, phase difference $\Delta=\pi/5$). Left: $\theta$ changes the mean and the amplitude of the fringe. Right: $\phi$ moves the output along the fringe, which peaks at $\phi=-\Delta$. (e) A single intensity matrix allows many phases, but weights shared by all settings give a predicted phase up to the port phases; subtracting row 1 and column 1 gives the normalised phase $\varphi^{\rm norm}$, which is then tested on the chip.}
\label{fig:concept}
\end{figure}

The chip is a $12\times12$ Clements mesh of 66 MZIs arranged in 12 layers, A--L [Fig.~\ref{fig:concept}(a)]. Each MZI consists of two multimode-interference (MMI) couplers and two heaters [inset of Fig.~\ref{fig:concept}(a)]: the internal phase $\theta$, between the couplers, sets how the light is split, and the external phase $\phi$, on one input, sets the phase of that input relative to the other. Following Ref.~\cite{Cataldo2026PRL}, we describe the chip with a physics-informed digital twin, a model with the same mesh in which every MZI is a $2\times2$ matrix set by its two phases, the imbalances $\epsilon_{1,2}$ of its couplers and its transmission $\tau$. The product of these matrices is the complex transfer matrix $\U^{\rm twin}(\Iv)$ of the mesh for the heater currents $\Iv$. Each heater converts electrical power into phase,
\begin{equation}
  \theta(I) = g\,K\,I^2 + b ,
  \label{eq:current2phase}
\end{equation}
where $KI^2=2\pi R(I)I^2/P_{2\pi}$ converts the electrical power $R(I)I^2$ of the heater into phase, with the resistance $R(I)$ and the power $P_{2\pi}$ for a $2\pi$ shift measured beforehand (Methods), and the gain $g>0$ and the offset $b$ are weights of the heater. Eq.~\eqref{eq:current2phase} is written for $\theta$; $\phi$ follows the same law with its own weights.

All weights of the twin are physical quantities of the chip (Table~\ref{tab:weights}): the gain and offset of each heater, the imbalance of each coupler and the transmission of each MZI, together with the mixing of power between ports at the input and output of the chip and the efficiency of each detector. None of them changes with the currents.

\begin{table}[htbp]
\centering
\caption{Weights of the twin, learned from the measured powers.}
\label{tab:weights}
\begin{tabular}{ll}
\hline
Symbol & Meaning \\
\hline
$g>0$ & gain of each heater \\
$b$ & offset of each heater \\
$\epsilon_{1,2}$ & imbalance of each multimode-interference (MMI) coupler \\
$\tau$ & transmission of each MZI \\
$\bm M_{\rm in}$, $\bm M_{\rm out}$ & input and output mixing \\
$\eta$ & detector efficiencies \\
\hline
\end{tabular}
\end{table}

We train the twin on powers alone [Fig.~\ref{fig:concept}(b)]. For each of 4000 settings, all heater currents are set at random, light enters one input $k$ of the chip, and the 12 output powers $\bm p$ are recorded. The twin receives the same currents and input and predicts the powers $\hat{\bm p}$, given by $|U^{\rm twin}|^2$ after the input and output mixing and the detector efficiencies, and the weights are adjusted until $\hat{\bm p}$ matches $\bm p$ for the 2880 settings used for fitting (Methods). On 400 settings held out from training, the twin reproduces the measured powers with a Bhattacharyya fidelity $\FB=(\sum_j\sqrt{p_j\hat p_j})^2=0.9987$.

No phase is measured, yet the fitted weights constrain the phases. Consider first one MZI with light in one input [Fig.~\ref{fig:concept}(c); Supplementary Note~\ref{app:mzi}]. A measured power gives $\sin^2(\theta/2)$, and hence $\theta$ up to its sign and multiples of $2\pi$. With Eq.~\eqref{eq:current2phase}, each of these values is one linear equation for the two unknowns $g$ and $b$, that is, a line in the $(g,b)$ plane [Fig.~\ref{fig:concept}(c), left]. A sample at another current gives lines of another slope, and with a few samples the lines cross only near the true weights and their mirror $(-g,-b)$, which $g>0$ excludes [Fig.~\ref{fig:concept}(c), middle and right]. With $g$ and $b$ known, $\theta(I)$ is known at every current, and with it the complex matrix of the MZI, not only its power. The external phase $\phi$ cannot change the power when light enters one input of the MZI, because it then only shifts the phase of all the light. It can when coherent light reaches both inputs with a phase difference $\Delta$. The phase $\phi$ then adds to $\Delta$, and the output power follows a fringe in $\phi+\Delta$: $\theta$ sets the mean and the amplitude of the fringe, and the fringe peaks at $\phi=-\Delta$, where $\phi$ cancels the phase difference and the two inputs add in phase [Fig.~\ref{fig:concept}(d); Supplementary Eq.~\eqref{eq:mzi_two_inputs}]. Inside the mesh this happens even when light enters the chip through a single port, because the preceding MZIs split it, so the powers constrain the weights of $\phi$ as they do those of $\theta$. The exception is the first layer: its MZIs receive light in one input only, so their external phases never change a measured power, and they are not driven (Methods). In the mesh, a single setting still allows many phases, because different complex matrices have the same powers [Fig.~\ref{fig:concept}(e), left]. The weights, however, are shared by all settings. Each setting adds 11 independent powers, since its 12 normalised outputs sum to one, but no new weights, so the 2880 fitted settings give about $3.2\times10^4$ observations for 750 weights (Methods). This count alone does not prove that the weights are unique, so we test the consequence directly. With the fitted weights, the twin gives a complex transfer matrix $\U^{\rm twin}(\Iv)$ at every setting, including its phase [Fig.~\ref{fig:concept}(e)]; this phase is a prediction made without any phase measurement, and we test it on the chip separately from the intensities.

\subsection*{Normalised phase and phase-aware inverse design}

 Consider an $m\times m$ block of the mesh, such as the $6\times6$ block on ports 1--6 in Fig.~\ref{fig:concept}(a), and write its elements as $U_{ik}=|U_{ik}|\,\ee^{\ii\pp{ik}}$, where $k$ labels the input and $i$ the output of the block (the subscripted $\pp{ik}$ is a phase of the transfer matrix, not the external heater phase $\phi$). On its way to and from the block, the light picks up unknown phases, $\alpha_k$ before input $k$ and $\beta_i$ after output $i$ [$\bm{D}_{\rm in}$ and $\bm{D}_{\rm out}$ in Fig.~\ref{fig:concept}(a)], so the chip implements $\pp{ik}+\alpha_k+\beta_i$ instead of $\pp{ik}$, and a single phase cannot be measured. Since $\beta_i$ shifts every phase in row $i$ and $\alpha_k$ every phase in column $k$, subtracting row 1 and column 1 removes both [Fig.~\ref{fig:concept}(e), right]. The resulting combination
\begin{equation}
  \pn{ik}=\pp{ik}-\pp{i1}-\pp{1k}+\pp{11}
  \label{eq:norm_phase}
\end{equation}
does not depend on $\alpha_k$ and $\beta_i$ (Methods and Supplementary Note~\ref{sec:evaldetails}), and we call it the normalised phase. It is zero in row and column~1 by construction and sets how light from different inputs interferes inside the block. We display target, twin and measured phases through $\pn{ik}$ and quantify their agreement after fitting input and output phases (Methods).

To program a target unitary $\U^{\rm target}$ on such a block, only the MZIs of the block are tuned. The MZIs that couple the block to the rest of the mesh are set to the bar state (BAR), in which light stays in its waveguide, to suppress leakage; imperfect couplers can leave residual coupling. The current for $\theta=\pi$ is obtained by inverting Eq.~\eqref{eq:current2phase} with the trained weights. For given settings of the block heaters, the twin computes the full matrix $\U^{\rm twin}$, and the block is its $m\times m$ window
\begin{equation}
  W_{ik}=U^{\rm twin}_{p_0+i-1,\,p_0+k-1},\qquad
  \hat P_{ik}=\frac{|W_{ik}|^2}{\sum_{i'}|W_{i'k}|^2},
  \label{eq:window}
\end{equation}
where $p_0$ is the first port of the block and $\hat P_{ik}$ the predicted power of output $i$ for input $k$, normalised per input as on the chip. The block heaters are found by minimising
\begin{equation}
  \mathcal{L}=\mathcal{L}_{\rm int}\big(\hat P,|U^{\rm target}|^2\big)+\lambda_{\rm leak}\,\mathcal{L}_{\rm leak}
  +\lambda_{\rm ph}\,\Big\langle 1-\cos\!\big(\pnl{twin}{ik}-\pnl{target}{ik}\big)\Big\rangle ,
  \label{eq:loss}
\end{equation}
where $\mathcal{L}_{\rm int}$ compares the predicted and target intensities, $\mathcal{L}_{\rm leak}$ is the fraction of light that leaves the block, and the last term compares the normalised phases of the block and the target (Eq.~\eqref{eq:norm_phase}); $\lambda_{\rm leak}$ and $\lambda_{\rm ph}$ set the strength of these terms. With $\lambda_{\rm ph}=0$, the optimisation programs the intensities only: many settings reach the target intensities, each with its own phase [Fig.~\ref{fig:concept}(e), left], and which of them the optimisation reaches is set by its starting point. The optimisation starts from the analytic Clements decomposition of $\U^{\rm target}$~\cite{Clements2016}, whose phases are converted into currents with the trained weights through Eq.~\eqref{eq:current2phase}. From there, with the weights of the twin held fixed, a gradient-based optimiser (limited-memory Broyden--Fletcher--Goldfarb--Shanno, L-BFGS; Methods) adjusts the heater phases of the block. The final phases are converted into currents once more with Eq.~\eqref{eq:current2phase} and applied to the chip. At the solved currents $\Iv^\ast$, the twin predicts both the intensities and the normalised phase $\pnl{twin}{ik}(\Iv^\ast)$ of the programmed block.

\subsection*{On-chip phase measurement}

\begin{figure}[t]
\centering
\includegraphics[width=\textwidth]{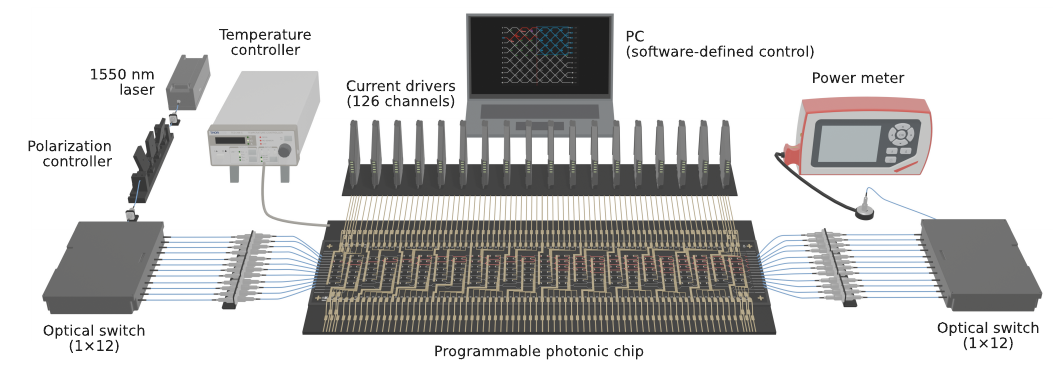}
\caption{Experimental setup. Light from a 1550\,nm laser passes a polarisation controller and an input $1\times12$ optical switch, which selects one of the 12 fibres of a fibre array and thus the input port of the programmable photonic chip. An output fibre array and a second $1\times12$ optical switch route one output port at a time to the power meter. Current drivers set the heater currents through the flex cable, a temperature controller stabilises the chip, and a PC connected to the switches, drivers and power meter runs the software-defined control.}
\label{fig:setup}
\end{figure}

\begin{figure}[htbp]
\centering
\includegraphics[width=\textwidth]{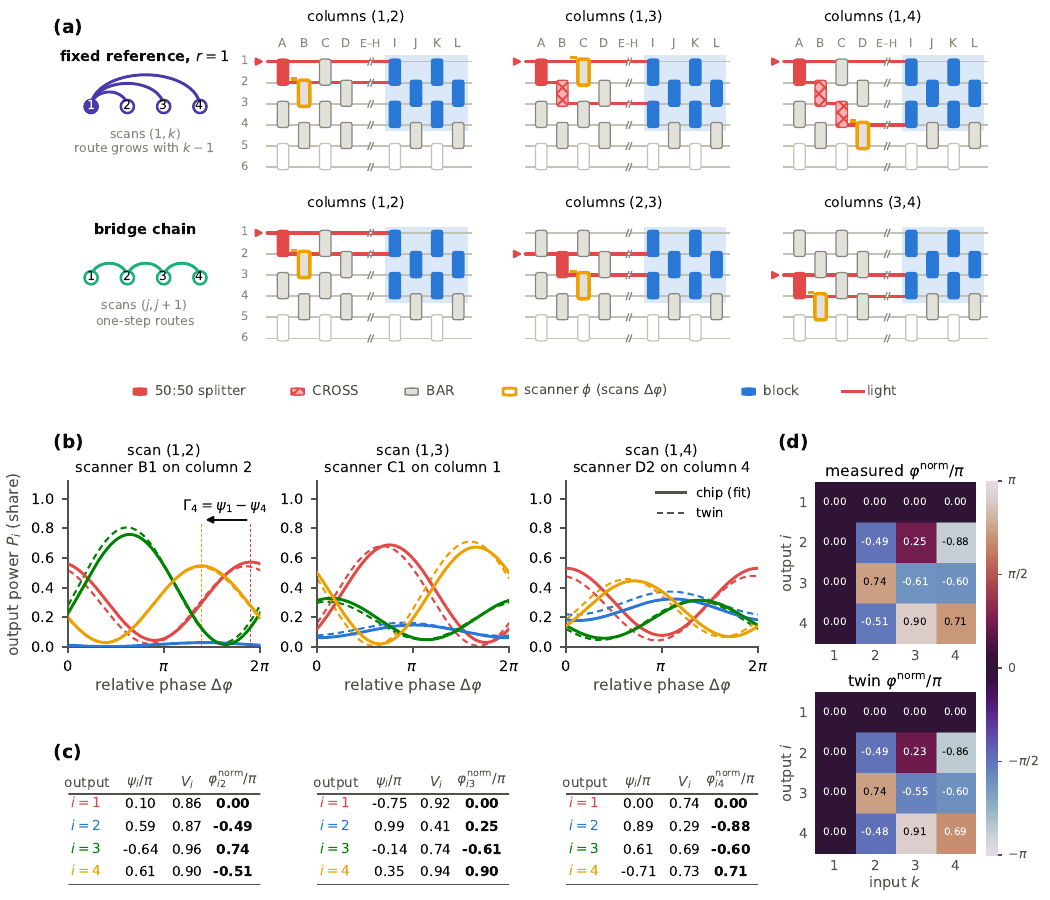}
\caption{On-chip phase measurement of a $4\times4$ block on ports 1--4 of the $12\times12$ mesh (layers I--L; ports 1--6 shown, BAR layers E--H cut out). (a) The $m-1=3$ scans of each protocol and their scan graphs. Top, fixed reference $r=1$: the route to column $k$ grows with $k-1$. Bottom, bridge chain: each scan couples neighbouring columns. (b) Fringes of the three fixed-reference scans, plotted against the relative phase $\Delta\varphi$ of column~1 with respect to column~$k$. When the scanner sits on column~$k$ (scans $(1,2)$ and $(1,4)$), its sweep is reversed so that all three scans share this axis. Solid lines show cosine fits to the measured forward and backward sweeps, and dashed lines show the twin's simulated fringes; the arrow marks $\Gamma_4=\psi_1-\psi_4$ for output 4. (c) Fitted fringe phases $\psi_i$, visibilities $V_i$ and the resulting normalised phase $\varphi^{\rm norm}_{ik}=\psi_1-\psi_i$, in units of $\pi$. (d) Measured normalised phase and the phase directly predicted from the twin's block transfer matrix for the programmed currents, in units of $\pi$.}
\label{fig:protocols}
\end{figure}

The chip is operated with the setup of Fig.~\ref{fig:setup}. Light from a laser at 1550\,nm passes a polarisation controller and a $1\times12$ optical switch that selects the input port of the chip; a second $1\times12$ switch selects the output port read by a power meter. Current drivers set the heater currents, and a temperature controller stabilises the chip (device and instruments in Supplementary Note~\ref{sec:chipnote}). To measure $\pn{ik}$ of a block, we generate two coherent beams on the chip and fit their interference fringes (Fig.~\ref{fig:protocols}; equations in Methods). One MZI in front of the block is set to 50:50 and splits light from a single chip input into two beams, which enter inputs $j$ and $k$ of the block and thus probe columns $j$ and $k$ of its matrix. Where these inputs are not adjacent, MZIs in the cross state (CROSS), which moves light to the neighbouring waveguide, route one beam to the required input. The external phase shifter of a following MZI on one path serves as the scanner. We step its current, convert the current into a phase $\Dphi$ with Eq.~\eqref{eq:current2phase}, record the output powers and fit the fringes with Eq.~\eqref{eq:fringe}. The unknown phase that the two beams pick up before the block is the same at every output of a scan, so subtracting the fringe phase at output $i$ from that at output~1 removes it. The $m-1$ scans are combined through Eq.~\eqref{eq:protocols}, either against one input kept as reference (the fixed-reference protocol) or along a chain of neighbouring inputs (the bridge chain). The measurement uses the twin to configure the routes and calibrate the scanner, but does not insert the predicted block phase into the fringe reconstruction. Figure~\ref{fig:protocols}(b)--(d) shows a fixed-reference measurement of a $4\times4$ block on ports 1--4 with reference column $r=1$. The measured fringes yield the normalised phase of the block, which agrees with the phase directly predicted by the twin with a mean absolute deviation of $0.065$\,rad and a maximum deviation of $0.19$\,rad [Fig.~\ref{fig:protocols}(d)].

\subsection*{Programming the phase from a random start}

The optimisation described above starts from a Clements decomposition of the target. This start already places the phases of the block MZIs close to the target, and an intensity-only optimisation from there stays close, so it cannot show whether the phase term itself programs the phase. We therefore start from a random setting of the block heaters, far from the target. From this start, the optimiser adjusts the heater phases step by step through the twin until the loss stops decreasing, once with $\lambda_{\rm ph}=0$ and once with $\lambda_{\rm ph}=0.5$. The twin and the start are the same for both values of $\lambda_{\rm ph}$, so the phase term is the only difference between the two programs (Methods). Starts far from the target also occur in practice when no accurate analytic start is available, e.g.\ for mesh architectures or imperfect networks that are programmed by numerical optimisation rather than by a direct decomposition~\cite{Perez2017,Burgwal2017,Fldzhyan2020,Saygin2020}, or when a complex-valued optical neural network is trained, since its network weights are updated iteratively from the previous setting rather than set from a decomposition~\cite{Hughes2018,Pai2023,Zhang2021CVNN}.

\begin{figure}[htbp]
\centering
\includegraphics[width=\textwidth]{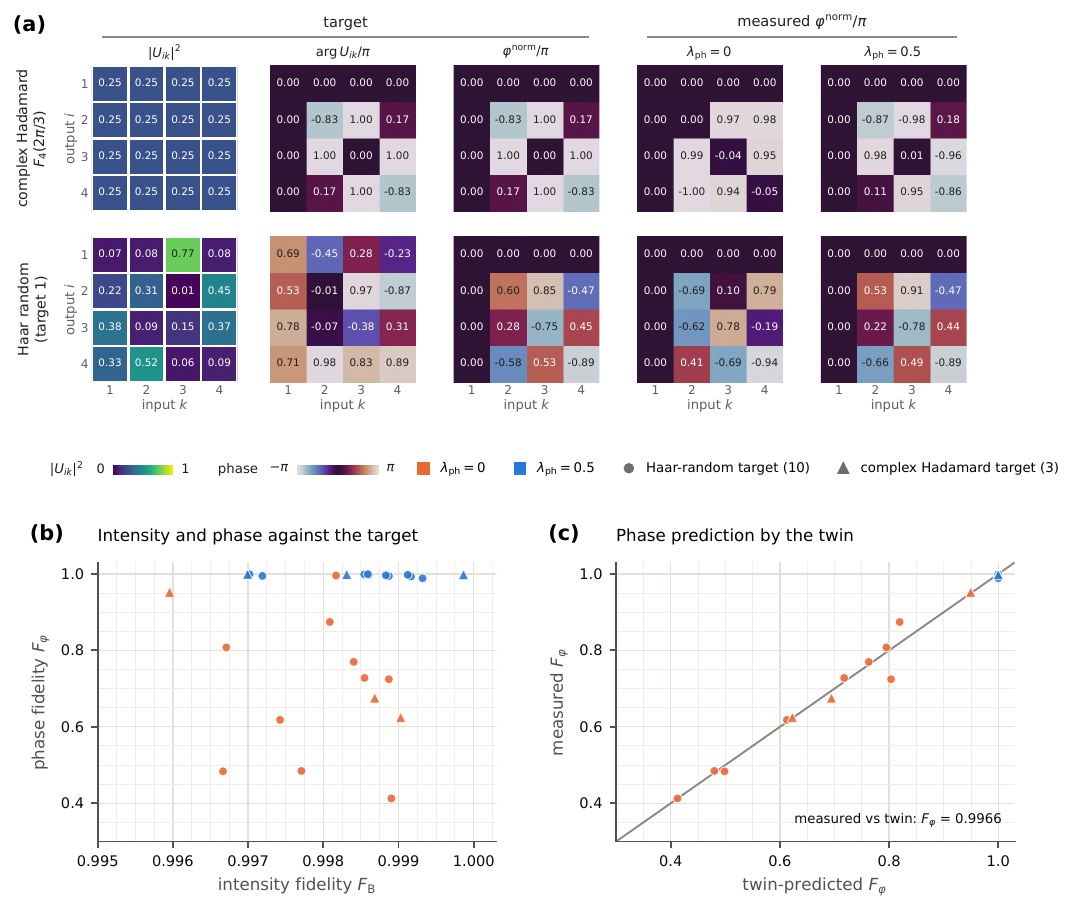}
\caption{Phase-aware programming from a random start. Ten Haar-random $4\times4$ targets and three complex Hadamard targets $F_4(a)$, $a=0,\pi/3,2\pi/3$, on ports 1--4 are each programmed with $\lambda_{\rm ph}=0$ and $0.5$ from a shared random start and measured with the bridge chain. (a) Two examples: the complex Hadamard target $F_4(2\pi/3)$, whose elements all have power $1/4$ (top), and Haar-random target~1 (bottom). From left: target power $|U_{ik}|^2$, target phase $\arg U_{ik}$, target normalised phase $\pn{ik}$, and the measured normalised phase of the programs with $\lambda_{\rm ph}=0$ and $0.5$. (b) Phase fidelity against intensity fidelity of every measured block, both against the target. (c) Measured phase fidelity against the phase fidelity of the block predicted by the twin at the same currents, both against the target; the line marks equality. Phase fidelities are evaluated after fitting and removing input and output port phases, on illuminated elements with fringe visibility $V\geq0.1$.}
\label{fig:randstart}
\end{figure}

We program ten Haar-random $4\times4$ targets on ports 1--4 in this way, together with three targets from the complex Hadamard family $F_4(a)$, $a=0,\pi/3,2\pi/3$~\cite{TadejZyczkowski2006} (Supplementary Note~\ref{sec:hadamard}), whose elements all have power $1/4$ for every $a$, so that intensities alone cannot tell them apart. Each of the 26 programs is applied to the chip and measured with the bridge chain, and the measured block is compared both with the target and with the block that the twin predicts at the same currents (Fig.~\ref{fig:randstart}). We score intensities with $\FB$ and phases with $\Fphi=|\langle\ee^{\ii\delta_{ik}}\rangle|^2$, where $\delta_{ik}$ is the wrapped phase residual after fitting input and output phases (Methods); $\Fphi=1$ means identical normalised phases.

Both programs reach the target intensities: for the Haar-random targets, the measured mean $\FB$ is $0.9979$ without and $0.9985$ with the phase term [Fig.~\ref{fig:randstart}(b)]. Their phases differ strongly. Without the phase term, the measured mean phase fidelity against the target is $\Fphi=0.690$, and only one of the ten programs lies close to the target phase; with the phase term, $\Fphi=0.996$, and the mean phase error after alignment falls from $0.41$ to $0.038$\,rad. The complex Hadamard targets show this in its clearest form. With identical target powers, the intensity-only programs reach $\FB=0.998$ but $\Fphi=0.676$, $0.953$ and $0.625$, whereas the phase-aware programs reach $\Fphi\geq0.998$ [Fig.~\ref{fig:randstart}(b)]. For $F_4(2\pi/3)$, the intensity-only program lands close to $F_4(3\pi/2)$, another member of the family with the same powers [Fig.~\ref{fig:randstart}(a)]. Programs with the same intensities can therefore implement very different phases, and the phase term fixes the phase without lowering the intensity fidelity. The measured phases follow the twin in both cases, including where they miss the target [Fig.~\ref{fig:randstart}(c)]. The mean measured--twin phase fidelity is $0.9961$ without and $0.9958$ with the phase term for the Haar-random targets, and $0.9992$ and $0.9987$ for the complex Hadamard targets. For the intensity-only programs, the twin's predicted phase fidelity against the target, $0.690$, equals the measured one, and the twin also singles out the one program that lands close to the target. The twin therefore predicts the phase that the chip implements, whether or not it is the target phase.

\subsection*{Block sizes and placements}

\begin{figure}[htbp]
\centering
\includegraphics[width=\textwidth]{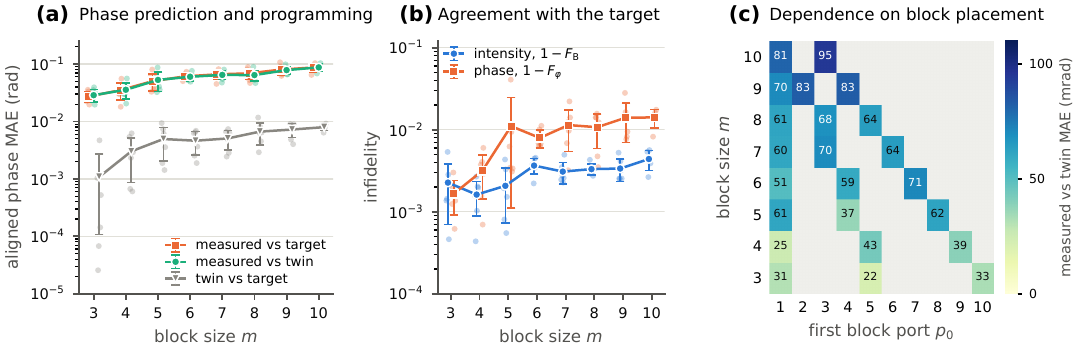}
\caption{Phase prediction across block sizes and placements. Blocks of size $m=3$--$10$ are programmed with $\lambda_{\rm ph}=0.5$ and measured with the bridge chain; two Haar-random targets per size give 46 target--placement configurations. (a) Aligned phase error of the measured block against the target and against the twin, and of the twin against the target. (b) Intensity and phase infidelity of the measured block against the target. Faint points, individual configurations; markers and error bars, mean and standard deviation. (c) Measured--twin phase error at each placement, averaged over the two targets; grey, not measured.}
\label{fig:experimental_sweep}
\end{figure}

To test the phase prediction beyond a single block, we program blocks of size $m=3$--$10$ at several positions on the mesh, with two Haar-random targets per size and 46 configurations in total, starting from a Clements decomposition and refining with $\lambda_{\rm ph}=0.5$ (Fig.~\ref{fig:experimental_sweep}). The measured--twin phase error grows from $0.029$\,rad at $m=3$ to $0.088$\,rad at $m=10$ [Fig.~\ref{fig:experimental_sweep}(a)], and at $m=10$ the measured block reaches mean intensity and phase fidelities of $0.9956$ and $0.9859$ against the target [Fig.~\ref{fig:experimental_sweep}(b)]. The agreement holds at every tested placement [Fig.~\ref{fig:experimental_sweep}(c)]. In two controls ($6\times6$ and $10\times10$), adding the phase term to a Clements warm start lowered the phase error predicted by the twin but not consistently the measured phase error, because the decomposition already places the block phases near the target (Supplementary Note~\ref{sec:clements_controls}). The phase term therefore matters most for programs that do not start from an accurate decomposition, as in Fig.~\ref{fig:randstart}.

\subsection*{Coherent inputs}

\begin{figure}[htbp]
\centering
\includegraphics[width=\textwidth]{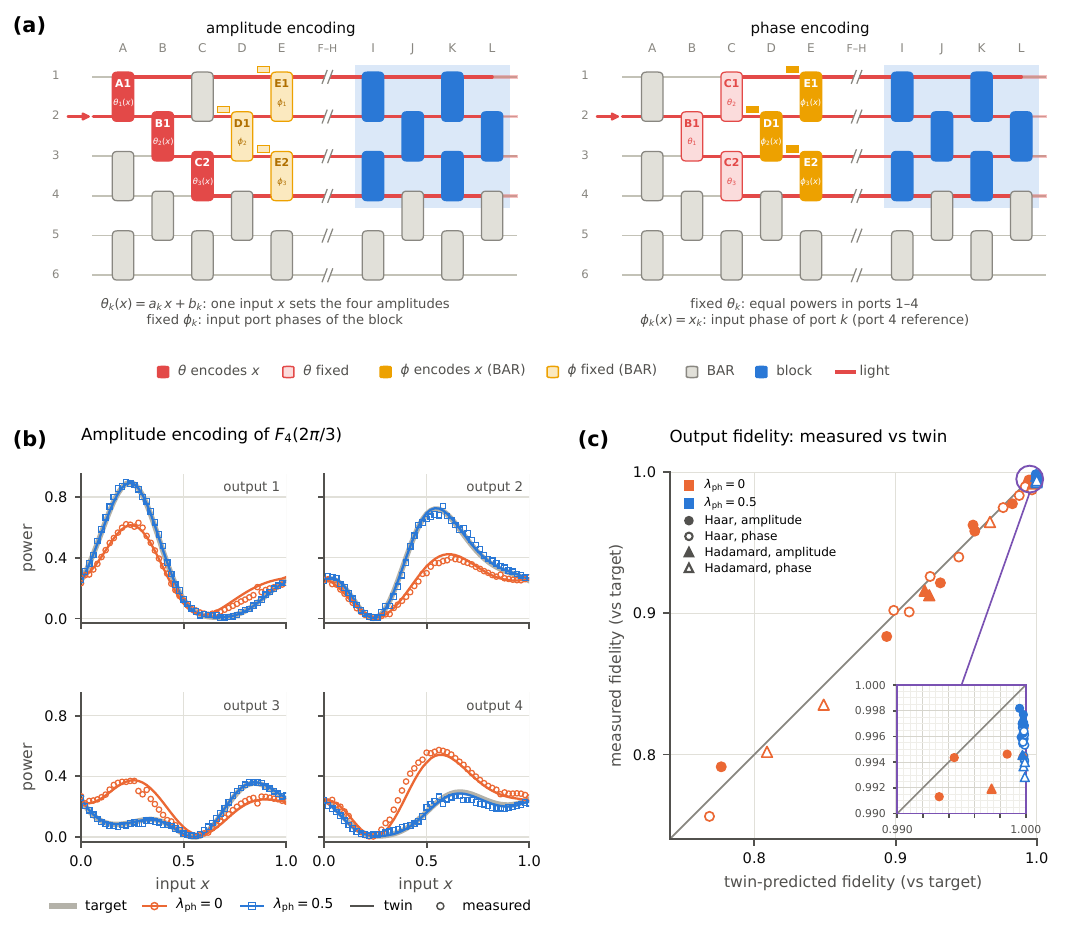}
\caption{Coherent inputs from on-chip angle encoding. (a) Amplitude and phase encoders (layers A--E) in front of a $4\times4$ block (blue); light enters port 2. Solid MZIs carry the input $x$; light MZIs are fixed. (b) Output powers of $F_4(2\pi/3)$ under amplitude encoding: target (grey), twin (lines) and measurement (markers) for $\lambda_{\rm ph}=0$ (orange) and 0.5 (blue). (c) Measured against twin-predicted output fidelity for each program and encoding of ten Haar-random (circles) and three complex Hadamard (triangles) targets; filled markers, amplitude encoding; open markers, phase encoding. Squares, random coherent inputs into $5\times5$ and $6\times6$ blocks with $\lambda_{\rm ph}=0.5$ (Supplementary Note~\ref{sec:coherent_large}). Inset: enlarged region near 1.}
\label{fig:encoding}
\end{figure}

In many applications, a programmed block receives coherent light in several inputs at once. Each output is then the interference of all input fields, $P_i\propto|\sum_k U_{ik}v_k|^2$, where $\bm v$ is the vector of input fields, and therefore depends on the phases of $\U$. We prepare such inputs on the chip with the MZIs in front of a $4\times4$ block, from light entering port 2 [Fig.~\ref{fig:encoding}(a)], and encode an input value into the phases of these MZIs in two ways (Table~\ref{tab:encoding}). In the amplitude encoding, a single number $x\in[0,1]$ sets the splitting of three MZIs, $\theta_k=k\pi x$, and with it the four real amplitudes of $\bm v$: as $x$ runs from 0 to 1, the input moves from a single port, through inputs that light all four ports, to another single port. In the phase encoding, three MZIs split the light equally into the four ports, and three numbers $x_1,x_2,x_3$ set the phases of ports 1--3 relative to port 4. The first encoding tests how the block mixes inputs of different amplitudes, the second how it responds to their phases. In both, the encoder heaters are solved through the twin so that the intended $\bm v$ reaches the block, with the block's input port phases compensated (Methods and Supplementary Note~\ref{sec:encoding}). This arrangement has the structure of one layer of a complex-valued photonic neural network~\cite{Zhang2021CVNN}: the encoder writes the input into the amplitudes and phases of $\bm v$, the block applies the complex matrix $\U$, and the photodiodes read $|\cdot|^2$.

\begin{table}[htbp]
\caption{On-chip angle encodings. Light enters port 2.}
\label{tab:encoding}
\centering
\setlength{\tabcolsep}{10pt}
\renewcommand{\arraystretch}{1.45}
\begin{tabular}{lll}
\hline
 & amplitude encoding & phase encoding \\
\hline
carries $x$ & A1: $\theta_1=\pi x$ & E1 (bar): $\phi_1=x_1$ \\
 & B1: $\theta_2=2\pi x$ & D1 (bar): $\phi_2=x_2$ \\
 & C2: $\theta_3=3\pi x$ & E2 (bar): $\phi_3=x_3$ \\
fixed & E1, D1, E2 (bar): $\phi_1,\phi_2,\phi_3$ & B1, C1, C2: $\theta=\pi/2$ (50:50) \\
 & (set once per program) & \\
\hline
input $\bm{v}$ & $v_1=\sin\frac{\theta_1}{2}$ & $v_1=\ee^{\ii x_1}/2$ \\
 & $v_2=\cos\frac{\theta_1}{2}\cos\frac{\theta_2}{2}$ & $v_2=\ee^{\ii x_2}/2$ \\
 & $v_3=\cos\frac{\theta_1}{2}\sin\frac{\theta_2}{2}\cos\frac{\theta_3}{2}$ & $v_3=\ee^{\ii x_3}/2$ \\
 & $v_4=\cos\frac{\theta_1}{2}\sin\frac{\theta_2}{2}\sin\frac{\theta_3}{2}$ & $v_4=1/2$ \\
samples & $x\in[0,1]$, 51 points & $x_1,x_2,x_3\in[0,2\pi)$, 30 random draws \\
\hline
output & $P_i\propto|(\U\bm{v})_i|^2$ & same \\
target & $Q_i\propto|(T\bm{v})_i|^2$ & same \\
fidelity & $F_{\rm out}=\big(\sum_i\sqrt{P_iQ_i}\big)^2$ & same \\
\hline
\end{tabular}
\end{table}

For the complex Hadamard target $F_4(2\pi/3)$, the intensity-only program ($\lambda_{\rm ph}=0$) produces outputs that deviate from the target ($F_{\rm out}=0.916$) but agree with the twin's prediction (measured--twin $F_{\rm out}=0.995$), whereas the phase-aware program follows the target ($F_{\rm out}=0.997$) [Fig.~\ref{fig:encoding}(b)].

The same holds across 13 targets, both encodings and 2106 inputs (51 values of $x$ and 30 random phase draws for each of the 26 programs) [Fig.~\ref{fig:encoding}(c)]. Intensity-only programs reach a mean $F_{\rm out}=0.936$, varying from $0.756$ to $0.997$ between programs, and the twin predicts each program's fidelity (Pearson $r=0.996$). Phase-aware programs reach $0.996$, with every program above $0.993$. For both, measured and predicted outputs agree with a mean fidelity of $0.996$. The same encoding extends to larger blocks: for random complex inputs prepared by all MZIs in front of $5\times5$ and $6\times6$ blocks, the measured outputs agree with the twin with a mean fidelity of $0.995$, whereas predictions that use only the powers of the target, with random input phases or an incoherent sum, reach $0.70$--$0.87$ (Supplementary Note~\ref{sec:coherent_large} and Supplementary Fig.~\ref{fig:coherent_large}).

\subsection*{Direction-of-arrival estimation}

\begin{figure}[t]
\centering
\includegraphics[width=\textwidth]{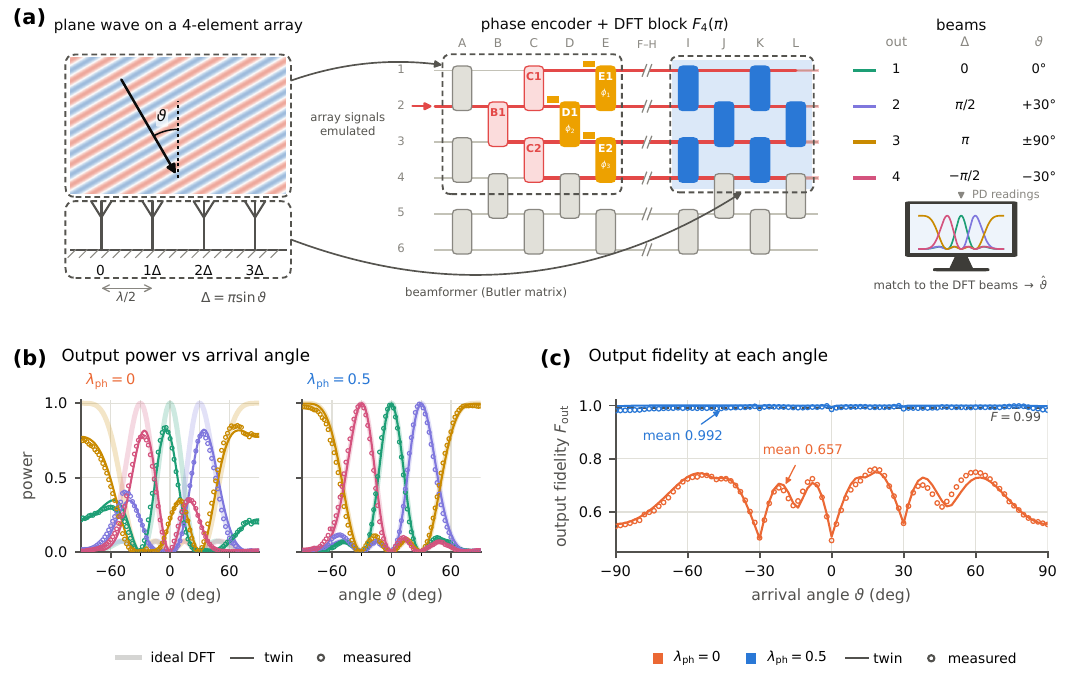}
\caption{Direction of arrival with an on-chip Butler matrix. (a) A plane wave at angle $\vartheta$ on a four-element array (spacing $\lambda/2$) gives element phases $0,\Delta,2\Delta,3\Delta$ with $\Delta=\pi\sin\vartheta$; these signals are emulated by the phase encoder (layers A--E), and the array's beamformer is the $4\times4$ block programmed as the DFT $F_4(\pi)$. Output $j$ is brightest for $\Delta=(j-1)\pi/2$; the angle $\hat\vartheta$ is obtained by matching the measured powers to the ideal DFT beams. Wavefronts are not to scale. (b) Output powers against $\vartheta$ for one program solved with $\lambda_{\rm ph}=0$ and $0.5$ from the same random start: ideal DFT (light), twin (lines) and measurement (markers). (c) Output fidelity against the ideal DFT beams at each of the 91 angles, averaged over five programs from different random starts; twin (lines) and measurement (markers). The dashed line marks $F_{\rm out}=0.99$.}
\label{fig:doa}
\end{figure}

As an application, we use a programmed block as an optical Butler matrix~\cite{ButlerLowe1961} for direction-of-arrival (DOA) estimation~\cite{KrimViberg1996} [Fig.~\ref{fig:doa}(a)]. Integrated meshes and photonic matrix multipliers have been used to find the direction of incoming optical beams~\cite{Milanizadeh2021,Zelaya2024}; here, the beams are set entirely by the phase of the programmed block, which the twin predicts and programs. A plane wave arriving at angle $\vartheta$ on a four-element array with half-wavelength spacing gives the element signals $v_k=\ee^{\ii(k-1)\Delta}/2$, $k=1,\ldots,4$, with $\Delta=\pi\sin\vartheta$. We emulate these signals with the phase encoder of Fig.~\ref{fig:encoding}: B1, C1 and C2 split the light equally, and D1, E1 and E2 write the element phases into ports 1--4. The block plays the role of the beamformer and is programmed as the four-point discrete Fourier transform (DFT), $F_4(\pi)$. Output $j$ adds the four signals in phase when $\Delta=(j-1)\pi/2$, so outputs 1--4 form beams at $\vartheta=0^\circ$, $+30^\circ$, $\pm90^\circ$ and $-30^\circ$. The angle is read out by matching the four measured powers to the ideal DFT beams (Supplementary Note~\ref{sec:doa}). Every element of the DFT has power $1/4$, as does every member of the family $F_4(a)$, so intensities alone do not fix the beams: they are determined by the phase of the DFT alone. We program the DFT from five random starts with $\lambda_{\rm ph}=0$ and $0.5$ and scan $\vartheta$ from $-90^\circ$ to $90^\circ$ in steps of $2^\circ$. With $\lambda_{\rm ph}=0.5$, the four beams peak at their design angles [Fig.~\ref{fig:doa}(b)]. With $\lambda_{\rm ph}=0$, the beams are distorted, and the measured outputs follow the twin's prediction rather than the DFT.

Across all 91 angles, the measured output fidelity against the ideal DFT beams averages $0.992$ for $\lambda_{\rm ph}=0.5$ and $0.657$ for $\lambda_{\rm ph}=0$, and the twin reproduces both curves [Fig.~\ref{fig:doa}(c)]. Within $|\vartheta|\leq60^\circ$, this gives a mean DOA error of $2.0^\circ$ with the phase term and $29^\circ$ without it, close to the twin's prediction of $28.5^\circ$ for the intensity-only programs. The phase term thus turns a block with the correct intensities into a working beamformer, and the twin predicts the outcome before measurement.

\section*{Discussion}

An intensity-trained physical model of a programmable mesh thus predicts more than intensities. The normalised phase of the programmed blocks agrees with the twin for intensity-only and phase-aware programs alike, including the specific wrong phases that intensity-only programs implement. Adding this phase to the inverse design programs it on the chip without phase measurements in the training data, and the difference becomes visible in every task with coherent inputs: the outputs of angle-encoded inputs, of random inputs into $5\times5$ and $6\times6$ blocks, and the beams of a Butler matrix follow the twin and, with the phase term, the target.

The remaining measured--twin difference, about $4\times10^{-3}$ in output infidelity, is 81 times larger than the repeatability of the measurement and is not explained by thermal crosstalk between heaters (Supplementary Note~\ref{sec:noise}); it reflects the accuracy of the model, which more training data or a richer description of the couplers and routing may improve. The benefit of the phase term depends on the starting point: from a Clements decomposition, which already places the block phases near the target, intensity-only refinement reached similar measured phase errors in the two tested controls. The on-chip protocols need free layers in front of the block, which limits the measured blocks to $10\times10$ for the bridge chain and $7\times7$ for the fixed reference on this processor, and the measurements cover ten Haar-random and three complex Hadamard targets in the paired comparison and two targets per size in the placement sweep.

Within this scope, the twin provides a phase-calibrated model of the chip that can be checked on the chip itself. This matters wherever a mesh is programmed by numerical optimisation rather than by an exact decomposition, for example for non-universal or imperfect architectures or during the training of optical neural networks, and wherever its outputs depend on interference between inputs, as in coherent signal processing and beamforming. For complex-valued photonic neural networks, this means that a model calibrated on output powers alone can set the phases of the network weights, not only their moduli.

\section*{Methods}

\subsection*{Digital twin: forward model and training}
\label{app:twin}

\emph{Forward model.} This section gives the full forward model with the weights of Table~\ref{tab:weights}. In Eq.~\eqref{eq:current2phase}, $K=(2\pi/P_{2\pi})\,R(I)$. Here $P_{2\pi}=2.46$\,mW is the electrical power for a $2\pi$ phase shift, obtained from a calibration of the chip in which the output power was recorded while the heater power was swept. The differences between heaters are absorbed by the gains, which are trained as $g=\ee^{\gamma}$ to keep them positive. The resistance $R(I)=R_0+R_2I^2$ of each heater is fitted to a current--voltage sweep of that heater before the optical measurements. To protect the heaters, the currents are limited to $1.3$\,mA. The external phase shifters of the first layer (A) are not driven and their weights are not used. Each MZI is
\begin{equation}
  T=\sqrt{\tau}\,C(\epsilon_2)\,\Phi(\theta)\,C(\epsilon_1)\,\Phi(\phi),\qquad
  C(\epsilon)=\begin{pmatrix} t & \ii\kappa\\ \ii\kappa & t\end{pmatrix},
  \label{eq:mzi_full}
\end{equation}
with $t,\kappa=\sqrt{(1\pm\tilde\epsilon)/2}$ and $\tilde\epsilon=0.95\tanh\epsilon$; for $\epsilon_{1,2}=0$ and $\tau=1$, it reduces to Supplementary Eq.~\eqref{eq:mzi_matrix}. The odd layers hold the MZIs on ports (1,2), (3,4), \dots, the even layers those on ports (2,3), (4,5), \dots, and $\U^{\rm twin}$ is the product of the 12 layers. For light sent to input $n$ of the setup, the predicted reading of detector $j$ is
\begin{equation}
  \hat p_j \propto \eta_j\sum_{i,k}(\bm M_{\rm out})_{ij}\,\big|U^{\rm twin}_{ik}\big|^2\,(\bm M_{\rm in})_{nk},
  \label{eq:readout}
\end{equation}
normalised to $\sum_j\hat p_j=1$. Here $k$ and $i$ run over the 12 input and output ports of the mesh, with the same order of indices as for the block elements $U_{ik}$; $(\bm M_{\rm in})_{nk}$ is the fraction of the input power that enters mesh input $k$ when input $n$ is selected, $(\bm M_{\rm out})_{ij}$ the fraction of the power leaving mesh output $i$ that reaches detector $j$, and $\eta_j$ the efficiency of detector $j$.

\emph{Data.} For each sample, the phase of every driven heater is drawn uniformly from $[0,2\pi)$ and converted into a current with Eq.~\eqref{eq:current2phase} at $g=1$, $b=0$; all currents stay below 1.27\,mA. Light enters one input port, chosen at random for each sample (308 to 372 samples per port), and the 12 output powers are recorded and normalised to unit sum.

\emph{Training.} The weights minimise
\begin{equation}
  \mathcal{L}_{\rm twin}=0.3\,\mathrm{MSE}(\bm p,\hat{\bm p})+0.1\,D_{\rm KL}(\bm p\,\|\,\hat{\bm p})+0.6\,\big(1-\FB\big)+\mathcal{R},
  \label{eq:loss_twin}
\end{equation}
where MSE is the mean squared error, $D_{\rm KL}$ the Kullback--Leibler divergence and $\FB$ the Bhattacharyya fidelity defined above, each averaged over the samples. The regularisation
\begin{equation}
\begin{aligned}
  \mathcal{R}={}&6\times10^{-4}\langle\epsilon^2\rangle+4\times10^{-4}\langle(g-1)^2\rangle\\
  &+2\times10^{-3}\big(\langle(\bm M_{\rm in}-\openone)^2\rangle+\langle(\bm M_{\rm out}-\openone)^2\rangle\big)
\end{aligned}
  \label{eq:reg}
\end{equation}
keeps the couplers, gains and mixings close to their nominal values. Of the 4000 samples, 400 are held out for testing and 720 for validation, and 2880 are used for training. We use AdamW (learning rate $10^{-3}$, weight decay $10^{-5}$, batch size 64), halve the learning rate when the validation fidelity stops improving for 10 epochs, and stop after 50 epochs without improvement. The weights with the highest validation fidelity are kept. The training starts from the weights of an earlier twin of the same chip, which was also trained on intensities alone. The twin has 750 weights that affect a measured power, while the 2880 training samples give $2880\times11\approx3.2\times10^4$ scalar observations after accounting for per-sample normalisation, since the 12 normalised outputs of each sample sum to one. On the test set, $\FB=0.9987$, between 0.9980 and 0.9992 for the 12 inputs. The trained gains lie within 1\,\% of the nominal value, and the coupler imbalances are $|\tilde\epsilon|\lesssim0.03$.

\subsection*{Phase-aware inverse design}
\label{app:inverse}

\emph{Variables.} Only the heaters of the block are solved. All other MZIs are held at fixed currents, the walls at $\theta=\pi$, obtained by inverting Eq.~\eqref{eq:current2phase}. The variables are the heater phases of the block, not the currents: a phase is periodic and unbounded, so a heater near $0$ or $2\pi$ is not pinned at a current limit, and the gradient does not vanish at $I=0$, where $\mathrm{d}\theta/\mathrm{d}I=0$. After the optimisation, each phase is reduced modulo $2\pi$ and converted into a current by inverting Eq.~\eqref{eq:current2phase}, which is quadratic in $I^2$ because of $R(I)$; a full $2\pi$ fits below the current limit.

\emph{Loss.} In Eq.~\eqref{eq:loss},
\begin{equation}
  \mathcal{L}_{\rm int}=\tfrac12\,\mathrm{MSE}\big(\hat P,|U^{\rm target}|^2\big)+\tfrac12\,(1-\bar B),
  \qquad
  \bar B=\Big\langle\sum_i\big(\hat P_{ik}\,|U^{\rm target}_{ik}|^2\big)^{1/2}\Big\rangle_k ,
  \label{eq:Lint}
\end{equation}
where $\bar B$ is the Bhattacharyya coefficient averaged over the inputs $k$. The leakage is
\begin{equation}
  \mathcal{L}_{\rm leak}=\Big\langle 1-\frac{\sum_{i\in{\rm block}}|U^{\rm twin}_{ik}|^2}{\sum_i|U^{\rm twin}_{ik}|^2}\Big\rangle_k ,
  \label{eq:Lleak}
\end{equation}
and $\lambda_{\rm leak}=0.5$. For the phase term, $\pn{ik}$ of the column-normalised window and of the target are computed with Eq.~\eqref{eq:norm_phase}. The cost of a cell is $1-\cos(\pnl{twin}{ik}-\pnl{target}{ik})$, which we evaluate from the complex elements without computing an angle, so that it has no jump at $\pm\pi$. The informative inner cells are those for which the four target elements $U^{\rm target}_{ik}$, $U^{\rm target}_{i1}$, $U^{\rm target}_{1k}$ and $U^{\rm target}_{11}$ all have squared magnitude greater than $10^{-3}$. How the first row and column enter the normalisation of this term is described in Supplementary Note~\ref{sec:evaldetails}. If there is no informative inner cell, as for a diagonal target, the phase term vanishes.

\emph{Optimiser.} Unless random initialisation is specified below, the optimisation starts from the analytic Clements decomposition of $\U^{\rm target}$~\cite{Clements2016} on the block MZIs, whose phases are converted into currents with Eq.~\eqref{eq:current2phase}. This decomposition assumes ideal couplers; the subsequent optimisation includes the twin's fitted coupler imbalances. From there we run L-BFGS with a memory of 20 steps and a backtracking line search, through the differentiable forward model. It stops when the loss changes by less than $10^{-11}$, the gradient falls below $10^{-7}$, or no step lowers the loss, with at most 300 iterations. If L-BFGS does not improve on the warm start, the warm start is kept. For the $4\times4$ block of Fig.~\ref{fig:protocols} with $\lambda_{\rm ph}=0.5$, the optimisation takes 13 iterations and about 1\,s.

\subsection*{Normalised phase and on-chip measurement protocols}
\label{app:protocols}

\emph{Port phases.} A photodiode measures $|U_{ik}|^2$ but not $\pp{ik}$, and the light picks up unknown phases outside the block: $\alpha_k$ before input $k$, from fibres, waveguides and the MZIs in front of the block, which changes a measured power only when light enters several ports at once, and $\beta_i$ after output $i$, on its way to the detector, which does not change any measured power. Each phase belongs to one port, so the chip implements
\begin{equation}
  \tilde{\U}=\bm{D}_{\rm out}\,\U\,\bm{D}_{\rm in},\qquad
  \bm{D}_{\rm in}=\mathrm{diag}(\ee^{\ii\alpha_k}),\qquad
  \bm{D}_{\rm out}=\mathrm{diag}(\ee^{\ii\beta_i}),
  \label{eq:gauge}
\end{equation}
or, for each element,
\begin{equation}
  \tilde U_{ik}=|U_{ik}|\,\ee^{\ii\tilde\varphi_{ik}},\qquad
  \tilde\varphi_{ik}=\pp{ik}+\alpha_k+\beta_i .
  \label{eq:gauge_elem}
\end{equation}
Since $\alpha_k$ and $\beta_i$ are unknown, a single $\pp{ik}$ cannot be measured; only combinations in which they cancel can.

Subtracting from every phase the first element of its row and then that of its column removes both, and what remains is the normalised phase of Eq.~\eqref{eq:norm_phase} (step-by-step derivation in Supplementary Note~\ref{sec:evaldetails}).
The normalised phase measured on the chip is thus that of the block itself. It is defined wherever the four elements are non-zero. Any other row and column could serve as reference and would carry the same information; we use the first for simplicity.

\emph{Step 1: fringe.} Let light enter the block at inputs $j$ and $k$, and scan the phase $\Dphi$ of beam $j$. By Eq.~\eqref{eq:gauge_elem}, the field at output $i$ is
\begin{equation}
  E_i \propto |U_{ij}|\,\ee^{\ii(\tilde\varphi_{ij}+\Dphi)}
            + |U_{ik}|\,\ee^{\ii\tilde\varphi_{ik}},
  \label{eq:field}
\end{equation}
so its power oscillates with $\Dphi$ [Fig.~\ref{fig:protocols}(b)],
\begin{equation}
  P_i(\Dphi) \propto |U_{ij}|^2+|U_{ik}|^2+2|U_{ij}||U_{ik}|\cos\!\big(\Dphi+\fr{j,k}{i}\big),
  \label{eq:fringe}
\end{equation}
with the fringe phase
\begin{equation}
  \fr{j,k}{i}=\tilde\varphi_{ij}-\tilde\varphi_{ik}
  =\pp{ij}-\pp{ik}+\alpha_j-\alpha_k .
  \label{eq:fringe_phase}
\end{equation}
The output phase $\beta_i$ drops out. The input phases $\alpha_j$ and $\alpha_k$ now include the routes to the block and the unknown offset of the scanner, and change from scan to scan. A scanner on beam $k$ reverses the sign of $\fr{j,k}{i}$; unequal beam powers and detection efficiencies change only the fringe amplitude.

\emph{Step 2: remove the input phases with row~1.} Since $\alpha_j-\alpha_k$ arises before the block, it is the same at every output of a scan. Subtracting the fringe phase of output $i$ from that of output~1 of the same scan removes it:
\begin{equation}
  \Gam{j,k}{i}\equiv\fr{j,k}{1}-\fr{j,k}{i}=\big(\pp{ik}-\pp{1k}\big)-\big(\pp{ij}-\pp{1j}\big).
  \label{eq:gamma}
\end{equation}

$\Gam{j,k}{i}$ is the phase of column $k$ relative to column $j$, referenced to row~1, and depends only on $\U$. For $j=1$ it is the normalised phase of Eq.~\eqref{eq:norm_phase},
\begin{equation}
  \Gam{1,k}{i} = \pp{ik}-\pp{i1}-\pp{1k}+\pp{11} = \pn{ik},
  \label{eq:gamma_norm}
\end{equation}
so a single scan of columns 1 and $k$ gives column $k$ of $\pn{}$.

\emph{Step 3: assemble the normalised phase.} Eq.~\eqref{eq:gamma} is the difference of a quantity of column $k$ and the same quantity of column $j$. $\Gamma$ is therefore antisymmetric, $\Gam{k,j}{i}=-\Gam{j,k}{i}$, and additive, $\Gam{j,k}{i}+\Gam{k,l}{i}=\Gam{j,l}{i}$, so $\Gam{1,k}{i}$ can be assembled from other pairs of columns. The \emph{fixed reference} keeps the laser on column $r$ and scans $(r,k)$ for every $k\neq r$; the \emph{bridge chain} scans only neighbouring columns:
\begin{equation}
  \pn{ik} = \Gam{r,k}{i}-\Gam{r,1}{i}, \qquad
  \pn{ik} = \sum_{j=1}^{k-1}\Gam{j,j+1}{i}.
  \label{eq:protocols}
\end{equation}
The errors of the two protocols and the checks between them are discussed in Supplementary Note~\ref{sec:evaldetails} and compared in Supplementary Table~\ref{tab:methods}. The amplitudes follow from lighting one column alone, $|U_{ik}|^2=P_{ik}/\sum_{i'}P_{i'k}$.

\emph{Step 4: build the beams on chip.} As shown in Fig.~\ref{fig:protocols}(a), the two beams share the input fibre, so drifts in front of the chip cancel. The routes must fit in front of the block, which limits the bridge chain to $m=N-2$ and the fixed reference to $m=\lfloor(2N-1)/3\rfloor$ (Supplementary Table~\ref{tab:methods}); larger blocks, including the full chip, require external dual input. The fixed reference needs longer routes. In the simulated measurements, the bridge chain gives smaller mean errors despite combining more scans for the later columns (Supplementary Note~\ref{sec:numerics}).

\subsection*{Initialisation and evaluation}
\label{sec:evaluation}

\emph{Haar-random and complex Hadamard targets of Fig.~\ref{fig:randstart}.} The ten Haar-random $4\times4$ targets are the first ten targets drawn with fixed seeds. Before any chip measurement, each target is screened in the twin: the $\lambda_{\rm ph}=0$ program must reach an intensity fidelity of at least $0.999$, and the $\lambda_{\rm ph}=0.5$ program an intensity fidelity of at least $0.999$ and a phase fidelity of at least $0.99$ against the target. All ten targets passed, so none was excluded. For each target, five random initial settings are generated, the one giving the lowest final loss with $\lambda_{\rm ph}=0.5$ is selected, and both $\lambda_{\rm ph}=0$ and $0.5$ are optimised from this same setting. The complex Hadamard targets $F_4(a)$, $a=0,\pi/3,2\pi/3$, and the DFT targets of Fig.~\ref{fig:doa} are programmed in the same way (Supplementary Note~\ref{sec:hadamard}).

The numerical sweep of $\lambda_{\rm ph}$ in Supplementary Fig.~\ref{fig:numerics}(b) follows the same procedure, with the start selection described in Supplementary Note~\ref{sec:numerics}. For the size and placement sweep in Fig.~\ref{fig:experimental_sweep}, we instead use a Clements decomposition as the warm start. Its phases are converted into currents using the twin's calibrated parameters, followed by L-BFGS optimisation through the twin with $\lambda_{\rm ph}=0.5$.

\emph{Intensity fidelity.} For two non-negative power matrices $A$ and $B$, each input column is normalised separately within the measured block:
\begin{equation}
  P_{ik}=\frac{A_{ik}}{\sum_{j=1}^{m}A_{jk}},\qquad
  Q_{ik}=\frac{B_{ik}}{\sum_{j=1}^{m}B_{jk}}.
\end{equation}
The reported intensity fidelity is the mean squared Bhattacharyya coefficient over the $m$ inputs,
\begin{equation}
  \FB=\frac{1}{m}\sum_{k=1}^{m}\Big(\sum_{i=1}^{m}\sqrt{P_{ik}Q_{ik}}\Big)^2.
\end{equation}
For measured powers, the dark background is subtracted, negative values are clipped to zero, and the calibrated output efficiencies are removed before normalisation. The target powers are $|U^{\rm target}_{ik}|^2$. In measured--twin intensity comparisons, the twin predicts the detected powers with its calibrated output mixing included. Since the columns are normalised within the block, $\FB$ measures agreement of the output distributions and does not include loss or leakage outside the block.

\emph{Phase references.} We distinguish the phase directly predicted by the twin from the phase reconstructed from simulated scans. The direct prediction is obtained from the complex block transfer matrix at the programmed currents. Simulated measurements instead reproduce the dual-input scans and apply the same fringe-fitting and phase-reconstruction procedure as for the chip. The measured--twin phase comparisons in Figs.~\ref{fig:protocols}, \ref{fig:randstart} and \ref{fig:experimental_sweep} use the directly predicted block. Supplementary Fig.~\ref{fig:numerics}(a) compares the phase reconstructed from simulated scans with that directly predicted block.

\emph{Input and output phase alignment.} Let $X$ be the block being evaluated and $Y$ the reference block. We first form the element-wise phase difference
\begin{equation}
  d_{ik}=\arg(X_{ik}Y_{ik}^{*}).
\end{equation}
Unknown output and input phase differences are represented by $\hat\beta_i$ and $\hat\alpha_k$. They are fitted on a set of valid elements $\mathcal{A}$ by alternating updates that reduce the weighted circular loss
\begin{equation}
  J(\hat\alpha,\hat\beta)=\sum_{(i,k)\in\mathcal{A}}w_{ik}\left[1-\cos(d_{ik}-\hat\beta_i-\hat\alpha_k)\right],
\end{equation}
where $w_{ik}=|Y_{ik}|^2$. Because $J$ can have local minima when $X$ is far from $Y$, as for intensity-only programs from a random start, the fit is run from $\hat\beta_i=\hat\alpha_k=0$ and from 39 sets of random input phases $\hat\alpha_k$; each start performs 200 iterations of
\begin{equation}
  \hat\beta_i \leftarrow \arg\Big[\sum_{k:(i,k)\in\mathcal{A}}w_{ik}\,\ee^{\ii(d_{ik}-\hat\alpha_k)}\Big],\qquad
  \hat\alpha_k \leftarrow \arg\Big[\sum_{i:(i,k)\in\mathcal{A}}w_{ik}\,\ee^{\ii(d_{ik}-\hat\beta_i)}\Big].
\end{equation}
The aligned residual of the start with the highest phase fidelity on $\mathcal{A}$ is
\begin{equation}
  \delta_{ik}=\arg\big[\ee^{\ii(d_{ik}-\hat\beta_i-\hat\alpha_k)}\big].
\end{equation}
This fit removes input and output phase differences without choosing a single matrix element as the phase reference. For blocks close to the reference, all starts reach the same fit; for the 46 configurations of Fig.~\ref{fig:experimental_sweep}, the scores change by at most $4\times10^{-6}$ relative to a single start from zero.

\emph{Phase error and fidelity.} On the scoring set $\mathcal{S}$, containing $N_{\mathcal{S}}$ elements, we report the mean absolute wrapped phase error
\begin{equation}
  \bar\delta=\frac{1}{N_{\mathcal{S}}}\sum_{(i,k)\in\mathcal{S}}|\delta_{ik}|
\end{equation}
and the phase fidelity
\begin{equation}
  \Fphi=\Big|\frac{1}{N_{\mathcal{S}}}\sum_{(i,k)\in\mathcal{S}}\ee^{\ii\delta_{ik}}\Big|^2.
\end{equation}
Although the alignment is weighted by the reference intensities, both reported quantities are unweighted averages over the scoring elements. Phase infidelity is $1-\Fphi$. These aligned scores include valid illuminated elements in the first row and first column.

The normalised-phase maps, and the deviations reported for Fig.~\ref{fig:protocols}(d), use the convention given in Supplementary Note~\ref{sec:evaldetails}.

\emph{Valid elements.} A fitted fringe is written as $f(\Dphi)=f_0+f_c\cos\Dphi+f_s\sin\Dphi$, with visibility
\begin{equation}
  V=\frac{\sqrt{f_c^2+f_s^2}}{f_0}.
\end{equation}
A reconstructed phase is accepted for scoring only if every fringe needed for its reconstruction, at both the output being evaluated and the reference output, has $V\geq0.1$. For the bridge chain, this condition must hold for every bridge along the path to that column. Outputs with negligible fitted mean power are also rejected; the experimental analysis uses a minimum mean share of $10^{-4}$ of the block output, while the numerical scan implementation uses a numerical floor of $10^{-12}$. The first column is fixed by the phase convention and does not require a fringe of its own.

For the experimental aligned scores, the fitting and scoring sets are the same: reconstructed phases passing the visibility criterion and satisfying $|Y_{ik}|^2>10^{-3}$. For comparisons against the target, $Y=U^{\rm target}$; for comparisons against the directly predicted block, $Y=U^{\rm twin}_{\rm block}$. For model--target comparisons without scans, no fringe-visibility criterion is needed.

\emph{Averaging across configurations.} Each score is calculated separately for each programmed configuration; how configurations are averaged in each figure is described in Supplementary Note~\ref{sec:evaldetails}.

\subsection*{On-chip encoding of coherent inputs}
\label{sec:methods_encoding}
Light enters chip port 2, and the MZIs in layers A--E in front of the $4\times4$ block (layers I--L) prepare the input vector $\bm v$ (Table~\ref{tab:encoding}); the remaining MZIs between the encoder and the block are set to BAR. The encoder does not use the programmed block. For each program, the twin provides one fixed calibration: the input port phases $\bm D_{\rm in}$ of the programmed block, obtained by fitting $\bm D_{\rm out}\,\U^{\rm target}\,\bm D_{\rm in}$ to the twin's block, and the column transmissions $\bm N$ of the block. For each input, the six encoder heaters are solved through the twin with the block set to BAR, so that the field reaching the block is proportional to $\bm N^{-1}\bm D_{\rm in}^{\ast}\bm v$, from three starts (two for the DOA scan), the first being the solution for the previous input. The program is then restored with the encoder currents unchanged. At regular intervals, the encoder alone is measured with the block at BAR, and a fixed reference configuration is measured to monitor drift (Supplementary Note~\ref{sec:encoding}).

The measurement order interleaves the programs: for each input, every program of a target is set and read in turn, so slow drifts affect the programs with $\lambda_{\rm ph}=0$ and $0.5$ alike. The four outputs of the block are read with the dark background subtracted and the calibrated detector efficiencies removed, and normalised to unit sum. The output fidelity is $F_{\rm out}=(\sum_i\sqrt{P_iQ_i})^2$ (Table~\ref{tab:encoding}), averaged over the inputs of a program.

\subsection*{Direction-of-arrival measurement}
\label{sec:methods_doa}
For the DOA experiment, the phase encoder writes the element phases $(k-1)\Delta$, $\Delta=\pi\sin\vartheta$, relative to port 4. The arrival angle is scanned from $-90^\circ$ to $90^\circ$ in steps of $2^\circ$ (91 angles) for each program. The four normalised output powers are compared with the ideal DFT beams $Q_j(\vartheta)=|\sum_k(F_4(\pi))_{jk}v_k(\vartheta)|^2$, computed on a $0.25^\circ$ grid, and the estimate $\hat\vartheta$ is the grid angle with the smallest squared difference. The DOA error is averaged over $|\vartheta|\leq60^\circ$, because near $\pm90^\circ$ the DFT beams cannot distinguish the two sides of the array ($\Delta=\pm\pi$). The DFT is programmed from five random starts; three targets $F_4(\pi+\delta)$ with $\delta=30^\circ$, $60^\circ$ and $90^\circ$ serve as controls (Supplementary Note~\ref{sec:doa}).

\subsection*{Data availability}
The data generated in this study are available from the corresponding authors upon reasonable request.

\subsection*{Code availability}
The code used to train the twin, solve the programs and analyse the measurements is not publicly available because it is part of ongoing development in the group. It is available from the corresponding authors upon reasonable request. The models, loss functions and algorithms are fully described in Methods.

\nocite{Cataldo2026Floquet,RausellCampo2026}
\bibliographystyle{naturemag}
\bibliography{references}

@article{Reck1994,
  author  = {Reck, M. and Zeilinger, A. and Bernstein, H. J. and Bertani, P.},
  title   = {Experimental realization of any discrete unitary operator},
  journal = {Phys. Rev. Lett.},
  volume  = {73},
  number  = {1},
  pages   = {58--61},
  year    = {1994},
  doi     = {10.1103/PhysRevLett.73.58}
}

@article{Clements2016,
  author  = {Clements, W. R. and Humphreys, P. C. and Metcalf, B. J. and Kolthammer, W. S. and Walmsley, I. A.},
  title   = {Optimal design for universal multiport interferometers},
  journal = {Optica},
  volume  = {3},
  number  = {12},
  pages   = {1460},
  year    = {2016},
  doi     = {10.1364/OPTICA.3.001460}
}

@article{Bogaerts2020,
  author  = {Bogaerts, W. and P{\'e}rez, D. and Capmany, J. and Miller, D. A. B. and Poon, J. and Englund, D. and Morichetti, F. and Melloni, A.},
  title   = {Programmable photonic circuits},
  journal = {Nature},
  volume  = {586},
  number  = {7828},
  pages   = {207--216},
  year    = {2020},
  doi     = {10.1038/s41586-020-2764-0}
}

@article{Shen2017,
  author  = {Shen, Y. and Harris, N. C. and Skirlo, S. and others},
  title   = {Deep learning with coherent nanophotonic circuits},
  journal = {Nat. Photonics},
  volume  = {11},
  number  = {7},
  pages   = {441--446},
  year    = {2017},
  doi     = {10.1038/nphoton.2017.93}
}

@article{Pai2023,
  author  = {Pai, S. and Sun, Z. and Hughes, T. W. and Park, T. and others},
  title   = {Experimentally realized in situ backpropagation for deep learning in photonic neural networks},
  journal = {Science},
  volume  = {380},
  number  = {6643},
  pages   = {398--404},
  year    = {2023},
  doi     = {10.1126/science.ade8450}
}

@article{Bandyopadhyay2021,
  author  = {Bandyopadhyay, S. and Hamerly, R. and Englund, D.},
  title   = {Hardware error correction for programmable photonics},
  journal = {Optica},
  volume  = {8},
  number  = {10},
  pages   = {1247--1255},
  year    = {2021},
  doi     = {10.1364/OPTICA.424052}
}

@article{Hamerly2022,
  author  = {Hamerly, R. and Bandyopadhyay, S. and Englund, D.},
  title   = {Accurate self-configuration of rectangular multiport interferometers},
  journal = {Phys. Rev. Appl.},
  volume  = {18},
  number  = {2},
  pages   = {024019},
  year    = {2022},
  doi     = {10.1103/PhysRevApplied.18.024019}
}

@article{Fyrillas2024,
  author  = {Fyrillas, A. and Faure, O. and Maring, N. and Senellart, J. and Belabas, N.},
  title   = {Scalable machine learning-assisted clear-box characterization for optimally controlled photonic circuits},
  journal = {Optica},
  volume  = {11},
  number  = {3},
  pages   = {427--436},
  year    = {2024},
  doi     = {10.1364/OPTICA.512148}
}

@article{Kuzmin2025,
  author  = {Kuzmin, S. and Dyakonov, I. and Straupe, S.},
  title   = {Leveraging a feature-based model for linear optical interferometer control},
  journal = {Phys. Rev. A},
  volume  = {112},
  number  = {5},
  pages   = {053515},
  year    = {2025},
  doi     = {10.1103/x8wd-83pt}
}

@article{Bantysh2024,
  author  = {Bantysh, B. I. and Chernyavskiy, A. Yu. and Fldzhyan, S. A. and Bogdanov, Yu. I.},
  title   = {Fast reconstruction of programmable interferometers with intensity-only measurements},
  journal = {Laser Phys. Lett.},
  volume  = {21},
  number  = {1},
  pages   = {015203},
  year    = {2024},
  doi     = {10.1088/1612-202X/ad0caf}
}

@misc{Laing2012,
  author        = {Laing, A. and O'Brien, J. L.},
  title         = {Super-stable tomography of any linear optical device},
  year          = {2012},
  note          = {arXiv:1208.2868}
}

@article{Dhand2016,
  author  = {Dhand, I. and Khalid, A. and Lu, H. and Sanders, B. C.},
  title   = {Accurate and precise characterization of linear optical interferometers},
  journal = {J. Opt.},
  volume  = {18},
  number  = {3},
  pages   = {035204},
  year    = {2016},
  doi     = {10.1088/2040-8978/18/3/035204}
}

@article{RahimiKeshari2013,
  author  = {Rahimi-Keshari, S. and Broome, M. A. and Fickler, R. and Fedrizzi, A. and Ralph, T. C. and White, A. G.},
  title   = {Direct characterization of linear-optical networks},
  journal = {Opt. Express},
  volume  = {21},
  number  = {11},
  pages   = {13450--13458},
  year    = {2013},
  doi     = {10.1364/OE.21.013450}
}

@article{Hoch2023,
  author  = {Hoch, F. and Giordani, T. and Spagnolo, N. and Crespi, A. and Osellame, R. and Sciarrino, F.},
  title   = {Characterization of multimode linear optical networks},
  journal = {Adv. Photonics Nexus},
  volume  = {2},
  number  = {1},
  pages   = {016007},
  year    = {2023},
  doi     = {10.1117/1.APN.2.1.016007}
}

@article{Cataldo2026PRL,
  title = {Interferometric Readout of Momentum-Space Topology in a Programmable Dissipative Photonic Circuit},
  author = {Cataldo, Andrea and Bergholtz, Emil J. and Leykam, Daniel and Gao, Jun and Elshaari, Ali W.},
  journal = {Phys. Rev. Lett.},
  volume = {137},
  issue = {20},
  pages = {206601},
  numpages = {6},
  year = {2026},
  month = {Nov},
  publisher = {American Physical Society},
  doi = {10.1103/xkht-64kc}
}

@article{Perez2017,
  author  = {P{\'e}rez, Daniel and Gasulla, Ivana and Crudgington, Lee and Thomson, David J. and Khokhar, Ali Z. and Li, Ke and Cao, Wei and Mashanovich, Goran Z. and Capmany, Jos{\'e}},
  title   = {Multipurpose silicon photonics signal processor core},
  journal = {Nat. Commun.},
  volume  = {8},
  number  = {1},
  pages   = {636},
  year    = {2017},
  doi     = {10.1038/s41467-017-00714-1}
}

@article{Burgwal2017,
  author  = {Burgwal, Roel and Clements, William R. and Smith, Devin H. and Gates, James C. and Kolthammer, W. Steven and Renema, Jelmer J. and Walmsley, Ian A.},
  title   = {Using an imperfect photonic network to implement random unitaries},
  journal = {Opt. Express},
  volume  = {25},
  number  = {23},
  pages   = {28236},
  year    = {2017},
  doi     = {10.1364/OE.25.028236}
}

@article{Fldzhyan2020,
  author  = {Fldzhyan, S. A. and Saygin, M. Yu. and Kulik, S. P.},
  title   = {Optimal design of error-tolerant reprogrammable multiport interferometers},
  journal = {Opt. Lett.},
  volume  = {45},
  number  = {9},
  pages   = {2632},
  year    = {2020},
  doi     = {10.1364/OL.385433}
}

@article{Saygin2020,
  author  = {Saygin, M. Yu. and Kondratyev, I. V. and Dyakonov, I. V. and Mironov, S. A. and Straupe, S. S. and Kulik, S. P.},
  title   = {Robust architecture for programmable universal unitaries},
  journal = {Phys. Rev. Lett.},
  volume  = {124},
  number  = {1},
  pages   = {010501},
  year    = {2020},
  doi     = {10.1103/PhysRevLett.124.010501}
}

@article{Hughes2018,
  author  = {Hughes, Tyler W. and Minkov, Momchil and Shi, Yu and Fan, Shanhui},
  title   = {Training of photonic neural networks through in situ backpropagation and gradient measurement},
  journal = {Optica},
  volume  = {5},
  number  = {7},
  pages   = {864},
  year    = {2018},
  doi     = {10.1364/OPTICA.5.000864}
}

@article{Kondratyev2026,
  author  = {Kondratyev, I. V. and Urusova, K. N. and Argenchiev, A. S. and Klushnikov, N. S. and Kuzmin, S. S. and Skryabin, N. N. and Golikov, A. D. and Kovalyuk, V. V. and Goltsman, G. N. and Dyakonov, I. V. and Straupe, S. S. and Kulik, S. P.},
  title   = {Effective programming of a photonic processor with complex interferometric structure},
  journal = {Phys. Rev. Appl.},
  volume  = {25},
  number  = {3},
  pages   = {034072},
  year    = {2026},
  doi     = {10.1103/btbd-w8fk}
}

@misc{RausellCampo2026,
  author        = {Rausell-Campo, J. R. and Melati, D. and Shastri, B. and P{\'e}rez-L{\'o}pez, D. and Capmany Francoy, J.},
  title         = {Universal neural network based calibration and control of programmable classical and quantum photonic integrated processors},
  year          = {2026},
  note          = {{arXiv}:2607.09301}
}

@article{TadejZyczkowski2006,
  author  = {Tadej, Wojciech and {\.Z}yczkowski, Karol},
  title   = {A concise guide to complex {H}adamard matrices},
  journal = {Open Syst. Inf. Dyn.},
  volume  = {13},
  pages   = {133--177},
  year    = {2006},
  doi     = {10.1007/s11080-006-8220-2}
}

@article{ButlerLowe1961,
  author  = {Butler, Jesse and Lowe, Ralph},
  title   = {Beam-forming matrix simplifies design of electronically scanned antennas},
  journal = {Electron. Des.},
  volume  = {9},
  pages   = {170--173},
  year    = {1961}
}

@article{KrimViberg1996,
  author  = {Krim, Hamid and Viberg, Mats},
  title   = {Two decades of array signal processing research: the parametric approach},
  journal = {IEEE Signal Process. Mag.},
  volume  = {13},
  number  = {4},
  pages   = {67--94},
  year    = {1996},
  doi     = {10.1109/79.526899}
}

@article{Milanizadeh2021,
  author  = {Milanizadeh, Maziyar and Toso, Fabio and Ferrari, Giorgio and Jonuzi, Tigers and Miller, David A. B. and Melloni, Andrea and Morichetti, Francesco},
  title   = {Coherent self-control of free-space optical beams with integrated silicon photonic meshes},
  journal = {Photonics Res.},
  volume  = {9},
  number  = {11},
  pages   = {2196},
  year    = {2021},
  doi     = {10.1364/PRJ.428680}
}

@article{Zelaya2024,
  author  = {Zelaya, Kevin and Miri, Mohammad-Ali},
  title   = {Photonic matrix multiplier makes a direction-finding sensor},
  journal = {Opt. Express},
  volume  = {33},
  pages   = {8396},
  year    = {2025},
  doi     = {10.1364/OE.542156}
}

@article{Tang2026,
  author  = {Tang, A. and Audhkhasi, R. and Tara, V. and Saxena, A. and Nath, G. and Majumdar, A.},
  title   = {Measurement of the complex scattering matrix in a nanoscale cavity array for boundary scattering tomography},
  journal = {Phys. Rev. Appl.},
  volume  = {26},
  number  = {5},
  pages   = {054010},
  year    = {2026},
  doi     = {10.1103/3ctw-4c89}
}

@misc{Cataldo2026Floquet,
      title={Floquet-driven light transport in programmable photonic processors via discretized evolution of synthetic magnetic fields}, 
      author={Andrea Cataldo and Rohan Yadgirkar and Ze-Sheng Xu and Govind Krishna and Ivan Khaymovich and Val Zwiller and Jun Gao and Ali W. Elshaari},
      year={2026},
      note={{arXiv}:2509.13184}
}

@article{Zhang2021CVNN,
  author  = {Zhang, H. and Gu, M. and Jiang, X. D. and Thompson, J. and Cai, H. and Paesani, S. and Santagati, R. and Laing, A. and Zhang, Y. and Yung, M. H. and Shi, Y. Z. and Muhammad, F. K. and Lo, G. Q. and Luo, X. S. and Dong, B. and Kwong, D. L. and Kwek, L. C. and Liu, A. Q.},
  title   = {An optical neural chip for implementing complex-valued neural network},
  journal = {Nat. Commun.},
  volume  = {12},
  pages   = {457},
  year    = {2021},
  doi     = {10.1038/s41467-020-20719-7}
}

@article{Xu2026Sync,
  author  = {Xu, Ze-Sheng and Cheng, Nan and Elmusrati, Mohammed S. and Yadgirkar, Rohan and Cataldo, Andrea and Wen, Rui and Krishna, Govind and Gao, Jun and Elshaari, Ali W.},
  title   = {Programmable non-{H}ermitian synchronization of light on a silicon photonic processor},
  journal = {ACS Photonics},
  volume  = {13},
  pages   = {6396--6404},
  year    = {2026},
  doi     = {10.1021/acsphotonics.6c01369}
}

\section*{Acknowledgements}
A.W.E. acknowledges support from the Knut and Alice Wallenberg (KAW) Foundation through the Wallenberg Centre for Quantum Technology (WACQT). J.G. acknowledges support from the Swedish Research Council (Ref: 2023-06671 and 2023-05288), Vinnova project (Ref: 2024-00466) and the G{\"o}ran Gustafsson Foundation. A.W.E. acknowledges support from the Swedish Research Council (VR) Starting Grant (Ref: 2016-03905). We thank Scott McHaffie for contributing to an earlier version of the intensity digital twin.

\section*{Author contributions}
H.L. developed the digital twin, the phase-aware inverse design and the on-chip protocols for phase measurement, performed the experiments and the analysis, and wrote the manuscript. A.C. and R.Y. contributed to methodology, experiments, manuscript, and discussion. J.G. and A.W.E. supervised the project. All authors discussed the results and contributed to the manuscript.

\section*{Competing interests}
The authors declare no competing interests.

\clearpage
\setcounter{section}{0}
\setcounter{figure}{0}
\setcounter{table}{0}
\setcounter{equation}{0}
\renewcommand{\thesection}{\arabic{section}}
\renewcommand{\thefigure}{\arabic{figure}}
\renewcommand{\thetable}{\arabic{table}}
\renewcommand{\theequation}{S\arabic{equation}}
\renewcommand{\theHsection}{S\arabic{section}}
\renewcommand{\theHfigure}{S\arabic{figure}}
\renewcommand{\theHtable}{S\arabic{table}}
\renewcommand{\theHequation}{S\arabic{equation}}
\makeatletter
\def\@seccntformat#1{\ifnum\pdfstrcmp{#1}{section}=0 Supplementary Note~\fi\csname the#1\endcsname\quad}
\makeatother
\renewcommand{\figurename}{Supplementary Fig.}
\renewcommand{\tablename}{Supplementary Table}

\begin{center}{\Large\bfseries Supplementary Information}\\[1ex]
{\large Unveiling gauge-invariant on-chip phases from light intensity}\\[1ex]
Huiwen Li, Andrea Cataldo, Rohan Yadgirkar, Jun Gao and Ali W. Elshaari\end{center}

\tableofcontents
\clearpage

\section{Single-MZI example: how power samples constrain the phase}
\label{app:mzi}

For a single Mach--Zehnder interferometer (MZI) with ideal couplers, $C=\tfrac{1}{\sqrt2}\begin{pmatrix}1&\ii\\ \ii&1\end{pmatrix}$, and phase shifters $\Phi(x)=\mathrm{diag}(\ee^{\ii x},1)$, the transfer matrix is
\begin{equation}
  T(\theta,\phi)=C\,\Phi(\theta)\,C\,\Phi(\phi)
  =\ii\,\ee^{\ii\theta/2}\begin{pmatrix}\sin\frac{\theta}{2}\,\ee^{\ii\phi} & \cos\frac{\theta}{2}\\[2pt] \cos\frac{\theta}{2}\,\ee^{\ii\phi} & -\sin\frac{\theta}{2}\end{pmatrix}.
  \label{eq:mzi_matrix}
\end{equation}
The internal phase $\theta$ sets the splitting ratio through $\sin(\theta/2)$ and $\cos(\theta/2)$, but it also adds the overall phase $\ee^{\ii\theta/2}$ and sets the signs of the elements, which affect the interference in the following MZIs. We take the bar output as an example. With light in one input, its power is
\begin{equation}
  P=|T_{11}|^2=\sin^2\!\frac{\theta(I)}{2},
  \label{eq:mzi_power}
\end{equation}
with $\theta(I)$ given by Eq.~\eqref{eq:current2phase}. Inverting Eq.~\eqref{eq:mzi_power} for a power $P_n$ measured at current $I_n$ gives
\begin{equation}
  \theta_n=\pm2\arcsin\sqrt{P_n}+2\pi\ell,\qquad \ell=0,\pm1,\pm2,\dots,
  \label{eq:theta_candidates}
\end{equation}
where $\ell$ is an integer that accounts for the $2\pi$ periodicity of the phase. One power thus corresponds to many values of $\theta$. The values $\theta$ and $-\theta$, for example, give the same power, $\sin^2(\theta/2)=\sin^2(-\theta/2)$, but different phases: the overall factor is $\ee^{\ii\theta/2}$ for one and $\ee^{-\ii\theta/2}$ for the other, and $\sin(-\theta/2)=-\sin(\theta/2)$ reverses the sign of the bar element. Both settings split the light in the same way but give it different phases. If we insert one of the values of Eq.~\eqref{eq:theta_candidates} into Eq.~\eqref{eq:current2phase}, only $g$ and $b$ are unknown, since the current $I_n$ is set in the experiment and $KI_n^2$ is a known number. This gives one equation for two unknowns, $\theta_n=g\,KI_n^2+b$, whose solutions form a straight line in the $(g,b)$ plane,
\begin{equation}
  b=s_n\,g+\theta_n ,\qquad s_n=-KI_n^2 ,
  \label{eq:line}
\end{equation}
with slope $s_n$. Since every value of Eq.~\eqref{eq:theta_candidates} gives such a line, a single sample allows a set of parallel lines [Fig.~\ref{fig:concept}(c), left]. A second sample at a different current $I_m$ gives another set of lines with a different slope $s_m$, and the true $(g,b)$ must lie on lines of both sets; with three samples, the sets of lines intersect only near a few points [Fig.~\ref{fig:concept}(c), middle]. Additional samples impose further constraints. In the noiseless example and parameter range shown in Fig.~\ref{fig:concept}(c), the misfit is concentrated around the generating point and its mirror $(-g,-b)$, up to shifts of $b$ by $2\pi$. This example illustrates how shared weights are constrained; it is not a general proof of uniqueness for a finite dataset or for the full mesh. The mirror point reverses $\theta$ at every current and gives the same powers; we take the solution with $g>0$, for which the phase increases with the heater power. Once $g$ and $b$ are known, $\theta(I)$ is known at every current, not only at the measured ones.

The external phase $\phi$ can affect the measured power when coherent light reaches both inputs of an MZI, as can occur inside the mesh even under single-input illumination of the chip. With fields $a_1$ and $a_2$ in the two inputs, the power in the bar output is
\begin{equation}
  P=\sin^2\!\frac{\theta}{2}\,|a_1|^2+\cos^2\!\frac{\theta}{2}\,|a_2|^2+\sin\theta\,|a_1||a_2|\cos(\phi+\Delta),
  \label{eq:mzi_two_inputs}
\end{equation}
where $\Delta$ is the phase difference between $a_1$ and $a_2$; for $a_2=0$ and $|a_1|=1$, Eq.~\eqref{eq:mzi_two_inputs} reduces to Eq.~\eqref{eq:mzi_power}. Both phases enter Eq.~\eqref{eq:mzi_two_inputs}, but in different ways: $\theta$ sets the mean and the amplitude of the interference, and $\phi$ only its position, with the maximum at $\phi=-\Delta$ [Fig.~\ref{fig:concept}(d)]. Varying the two heater currents independently provides samples with different dependence on $\theta$ and $\phi$. Just as $\theta$ enters the power through $\sin^2(\theta/2)$, $\phi$ enters through the interference term $\cos(\phi+\Delta)$, so a measured power $P_n$ again allows several values,
\begin{equation}
  \phi_n=\pm\arccos c_n-\Delta+2\pi\ell ,
  \label{eq:phi_candidates}
\end{equation}
where $c_n$ is the value of $\cos(\phi+\Delta)$ obtained from $P_n$ with Eq.~\eqref{eq:mzi_two_inputs} once $\theta$ and the input fields are known. With $\phi(I')=g'KI'^2+b'$ from Eq.~\eqref{eq:current2phase}, each value of Eq.~\eqref{eq:phi_candidates} gives one equation, $\phi_n=g'KI'^2_n+b'$, whose solutions form a straight line in the $(g',b')$ plane,
\begin{equation}
  b'=s'_n\,g'+\phi_n ,\qquad s'_n=-KI'^2_n ,
  \label{eq:line_phi}
\end{equation}
and multiple samples constrain $g'$ and $b'$ as in the single-heater example of Fig.~\ref{fig:concept}(c). In the mesh, $a_1$, $a_2$ and $\Delta$ depend on the preceding MZIs, so all weights are fitted together. The only exception is the first layer: since light enters the chip through one port at a time, its MZIs never receive light in both inputs, so their external phases never change a measured power. These phases only shift the phase of the light entering the chip, which no power measurement can see, and are not trained. With the weights of both heaters fixed, Eq.~\eqref{eq:mzi_matrix} gives the full complex matrix of every MZI, phases included.

\section{Numerical verification on the twin}
\label{sec:numerics}

\begin{figure}[htbp]
\centering
\includegraphics[width=\textwidth]{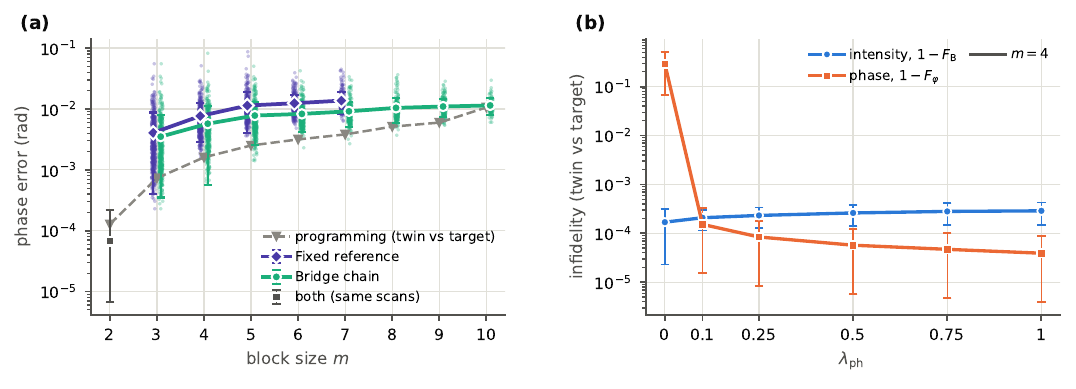}
\caption{Numerical verification on the $12\times12$ twin. (a) Phase error of the reconstructed block against the twin's block for both measurement protocols, using 50 Haar-random targets per block size at every allowed placement and $\lambda_{\rm ph}=0.5$. Points show individual target--placement configurations; markers and error bars show the mean and standard deviation. Grey shows the programming error of the twin's block against the target. (b) Intensity and phase infidelity of the programmed $4\times4$ block against the target versus $\lambda_{\rm ph}$, with a shared random start for each target across all values of $\lambda_{\rm ph}$. Results use 47 of 50 Haar-random targets on ports 1--4, retained using the intensity-fidelity criterion described in the text.}
\label{fig:numerics}
\end{figure}

Here we test the two steps of this work on the twin itself, which plays the role of the chip: the phase-aware programming of the Results and the on-chip measurement of Methods. For Supplementary Fig.~\ref{fig:numerics}(a), at each block size $m=2,\dots,10$, we draw 50 Haar-random targets and program each of them at every placement the mesh allows, with $\lambda_{\rm ph}=0.5$. We then emulate the dual-input scans on the twin, with 200 steps of the scanner and without added noise, and reconstruct the block phase as on the chip. The programming error compares the twin's block with the target, and the simulated measurement error compares the reconstructed block with the twin's block. Both errors are mean absolute wrapped phase differences after fitting and removing input and output phases. We also use the phase fidelity $\Fphi=|\langle\ee^{\ii\delta_{ik}}\rangle|^2$.

For Supplementary Fig.~\ref{fig:numerics}(b), we test 50 Haar-random $4\times4$ targets on ports 1--4. For each target, all values of $\lambda_{\rm ph}$ start from the same random setting of the block heaters. This setting is selected from five random starts by the final loss at $\lambda_{\rm ph}=1$. The selected initial setting is then shared by all tested values, $\lambda_{\rm ph}=0,0.1,0.25,0.5,0.75,1$. Three targets have an optimisation run with final intensity fidelity below $0.999$: target 2 at $\lambda_{\rm ph}=0$, target 16 at $0.5$, and target 24 at $0.75$. These targets are excluded at all values of $\lambda_{\rm ph}$, so every point uses the same 47 targets. The selection is based only on intensity fidelity. For these retained targets, the phase term improves the phase with little change in intensity fidelity. Without it, $1-\Fphi=0.29\pm0.22$; for every tested $\lambda_{\rm ph}\geq0.1$, the mean phase infidelity is below $2\times10^{-4}$, while the mean intensity infidelity stays between $1.7\times10^{-4}$ and $2.9\times10^{-4}$.

In Supplementary Fig.~\ref{fig:numerics}(a), the mean programming error increases from 0.7\,mrad for $m=3$ to 11\,mrad for $m=10$ (grey). In the simulated measurements, the bridge chain gives a mean phase error of 3.5\,mrad for $m=3$ and 11.4\,mrad for $m=10$. The fixed reference, limited to $m\leq7$, gives 4.1\,mrad for $m=3$ and 13.7\,mrad for $m=7$.

For Supplementary Fig.~\ref{fig:numerics}(a), the input and output phases are fitted using the reconstructed elements passing the original scan threshold $V\geq0.05$, with $|Y_{ik}|^2>10^{-3}$. The reported errors are then averaged over the subset passing $V\geq0.1$, without repeating the alignment. Between 1.5 and 7.8\,\% of the inner elements are excluded from scoring for $m\geq3$, including both low-intensity elements and those failing the visibility criterion.

\section{Normalised phase, protocol errors and evaluation details}
\label{sec:evaldetails}

This note collects details of the normalised phase, the on-chip protocols and the evaluation that are referred to in Methods.

\emph{Normalised phase.} The chip implements $\tilde\varphi_{ik}=\pp{ik}+\alpha_k+\beta_i$ [Eq.~\eqref{eq:gauge_elem}]. The unknown phases have a simple structure: $\beta_i$ adds the same constant to every phase in row $i$, and $\alpha_k$ adds the same constant to every phase in column $k$. Such constants are removed in two steps: we subtract from each row its first element, which removes $\beta_i$, and then from each column its first element, which removes $\alpha_k$,
\begin{align*}
&\begin{pmatrix}
\tilde\varphi_{11} & \cdots & \tilde\varphi_{1k} & \cdots\\
\vdots & & \vdots & \\
\tilde\varphi_{i1} & \cdots & \tilde\varphi_{ik} & \cdots\\
\vdots & & \vdots &
\end{pmatrix}
\xrightarrow{\text{rows}}
\begin{pmatrix}
0 & \cdots & \tilde\varphi_{1k}-\tilde\varphi_{11} & \cdots\\
\vdots & & \vdots & \\
0 & \cdots & \tilde\varphi_{ik}-\tilde\varphi_{i1} & \cdots\\
\vdots & & \vdots &
\end{pmatrix}\\
&\xrightarrow{\text{columns}}
\begin{pmatrix}
0 & \cdots & 0 & \cdots\\
\vdots & & \vdots & \\
0 & \cdots & \tilde\varphi_{ik}-\tilde\varphi_{i1}-\tilde\varphi_{1k}+\tilde\varphi_{11} & \cdots\\
\vdots & & \vdots &
\end{pmatrix}.
\end{align*}
What remains is the normalised phase of Eq.~\eqref{eq:norm_phase}. Inserting Eq.~\eqref{eq:gauge_elem},
\begin{align}
  \pn{ik} &= \tilde\varphi_{ik}-\tilde\varphi_{i1}-\tilde\varphi_{1k}+\tilde\varphi_{11} \nonumber\\
          &= (\pp{ik}+\alpha_k+\beta_i)-(\pp{i1}+\alpha_1+\beta_i)-(\pp{1k}+\alpha_k+\beta_1)+(\pp{11}+\alpha_1+\beta_1) \nonumber\\
          &= \pp{ik}-\pp{i1}-\pp{1k}+\pp{11}.
  \label{eq:norm_phase_steps}
\end{align}

\emph{Errors of the two protocols.} If each reconstructed scan difference has independent error with standard deviation $\sigma$, combining two scans gives $\sqrt{2}\sigma$ for a fixed reference with $r\neq1$ and $k\notin\{1,r\}$. With $r=1$, each non-reference column instead requires one scan and has error $\sigma$. The bridge chain gives $\sqrt{k-1}\sigma$ for column $k$ under the same independence assumption. Any closed loop of scans sums to zero, so where the protocols use different scans, e.g.\ (1,3) versus (1,2) and (2,3) in Fig.~\ref{fig:protocols}(a), they check each other.

\begin{table}[htbp]
\centering
\caption{The two on-chip protocols for an $m\times m$ block.}
\label{tab:methods}
\begin{tabular}{lcc}
\hline
 & Fixed reference & Bridge chain \\
\hline
Scans per block               & $m-1$ & $m-1$ \\
Scans per cell of column $k$  & 2 (1 for $r=1$) & $k-1$ \\
Longest route (ports)         & $\lceil(m-1)/2\rceil^{\ast}$ & 1 \\
$m_{\max}$                    & $\lfloor(2N-1)/3\rfloor$ & $N-2$ \\
$m_{\max}$ for $N=8$ / $12$   & 5 / 7 & 6 / 10 \\
\hline
\multicolumn{3}{l}{\footnotesize $^{\ast}$For a central reference column; the port parity of the block can force an off-centre $r$ and a longer route.} \\
\end{tabular}
\end{table}

\emph{Normalisation of the phase term.} In the phase term of Eq.~\eqref{eq:loss}, the cost is averaged over the informative inner cells defined in Methods. The implementation also includes illuminated first-row and first-column cells in the averaging denominator; their normalised-phase cost is zero by construction. Thus, these boundary cells set the normalisation of the phase term but add no phase constraint.

\emph{Normalised-phase maps.} The normalised-phase maps use a different convention:
\begin{equation}
  \pn{ik}=\arg\left(Y_{ik}Y_{11}Y_{i1}^{*}Y_{1k}^{*}\right),
\end{equation}
which sets the first row and first column to zero. The phase deviations reported for Fig.~\ref{fig:protocols}(d) compare these normalised phases directly, with wrapped differences, rather than using the fitted input and output phases of Methods. A normalised-phase comparison requires all four reference elements to have squared magnitude greater than $10^{-3}$. This four-element requirement is not applied to the aligned scores, which require only the individual reference element to satisfy $|Y_{ik}|^2>10^{-3}$.

\emph{Averaging across configurations.} Each fidelity or phase error is calculated separately for each programmed configuration. Figure~\ref{fig:randstart} shows every measured program, and the means quoted in the text are taken over the ten Haar-random or the three complex Hadamard targets at each value of $\lambda_{\rm ph}$. In Fig.~\ref{fig:experimental_sweep}, all 46 measured configurations are retained; the curves show the mean over the target--placement configurations at each size, and the placement map averages the two targets at each placement. Supplementary Fig.~\ref{fig:numerics}(a) likewise averages the scores of individual target--placement configurations. Where shown, error bars denote population standard deviations across configurations, rather than uncertainties on the mean. The 47-target selection for Supplementary Fig.~\ref{fig:numerics}(b) is described in Supplementary Note~\ref{sec:numerics}.

\section{Controls with a Clements warm start}
\label{sec:clements_controls}

We compare programming with $\lambda_{\rm ph}=0$ and $0.5$ for one $6\times6$ target on ports 1--6 and one $10\times10$ target on ports 1--10. For each target, both programs use the same Clements warm start, with phases converted into currents using the twin's calibrated weights, followed by limited-memory Broyden--Fletcher--Goldfarb--Shanno (L-BFGS) optimisation through the twin. The $\lambda_{\rm ph}=0$ programs therefore include intensity and leakage optimisation and are not measurements of the Clements initial settings alone. The two $6\times6$ programs are each measured using both the fixed-reference protocol, with reference column $r=3$, and the bridge chain. The two $10\times10$ programs are measured using the bridge chain. These six measurements correspond to two independent targets.

All measurements are analysed with a fringe-visibility threshold of $V\geq0.1$. The phase alignment, intensity threshold and fidelity definitions follow Methods. All 36 elements of each $6\times6$ block and all 100 elements of each $10\times10$ block pass the scoring criteria, including eligible elements in the first row and first column. The phase error predicted by the twin compares the directly predicted block with the target; the measured error compares the reconstructed block with the same target.

\begin{table}[htbp]
\centering
\caption{Programming controls from a Clements warm start. Both intensity fidelity $\FB$ and phase fidelity $\Fphi$ compare the measured block with the target. The twin and measured phase errors are aligned mean absolute wrapped phase errors against the target, in mrad. The two protocols for $m=6$ measure the same programmed blocks and do not represent additional independent targets.}
\label{tab:clements_controls}
\begin{tabular}{clccccc}
\hline
Block size $m$ & Protocol & $\lambda_{\rm ph}$ & $\FB$ & $\Fphi$ &
\shortstack{Twin phase\\error (mrad)} &
\shortstack{Measured phase\\error (mrad)} \\
\hline
6 & Fixed reference & 0   & 0.99713 & 0.99475 & 18.36 & 53.59 \\
6 & Fixed reference & 0.5 & 0.99650 & 0.99425 & 5.98  & 54.33 \\
6 & Bridge chain    & 0   & 0.99684 & 0.99708 & 18.36 & 40.05 \\
6 & Bridge chain    & 0.5 & 0.99659 & 0.99730 & 5.98  & 42.04 \\
10 & Bridge chain   & 0   & 0.99604 & 0.98023 & 79.94 & 90.25 \\
10 & Bridge chain   & 0.5 & 0.99624 & 0.98380 & 8.69  & 91.49 \\
\hline
\end{tabular}
\end{table}

For both targets, the phase term reduces the phase error predicted by the twin. The mean phase error falls from 18.36 to 5.98\,mrad for $m=6$ and from 79.94 to 8.69\,mrad for $m=10$. The measured mean phase errors, however, remain similar for the two values of $\lambda_{\rm ph}$, and the measured phase fidelity does not improve for every comparison. These controls show that, for the tested targets, programming from a Clements warm start can give good phase agreement even without the phase term. They do not establish statistical equivalence between the two values of $\lambda_{\rm ph}$ or identify the source of the remaining experimental discrepancy.

\section{Complex Hadamard targets}
\label{sec:hadamard}

The complex Hadamard matrices of order four form a one-parameter family~\cite{TadejZyczkowski2006},
\begin{equation}
  F_4(a)=\frac12\begin{pmatrix}
  1 & 1 & 1 & 1\\
  1 & \ii\ee^{\ii a} & -1 & -\ii\ee^{\ii a}\\
  1 & -1 & 1 & -1\\
  1 & -\ii\ee^{\ii a} & -1 & \ii\ee^{\ii a}
  \end{pmatrix},
  \label{eq:f4}
\end{equation}
which is unitary for every real $a$. Every element has power $|F_{4,ik}|^2=1/4$, so all members have the same intensity matrix and an intensity-only objective cannot distinguish them. They differ only in the normalised phase: by Eq.~\eqref{eq:norm_phase}, $\pn{22}=\pn{44}=a+\pi/2$ and $\pn{24}=\pn{42}=a-\pi/2$, while all other inner elements are $0$ or $\pi$ for every $a$. The four-point discrete Fourier transform (DFT) is $F_4(\pi)$, and $F_4(0)$ is the DFT with outputs 2 and 4 exchanged.

We read $a$ from a measured or predicted block $W$ as
\begin{equation}
  \hat a=\arg z,\qquad
  z=\tfrac14\Big[\ee^{\ii(\pn{22}-\pi/2)}+\ee^{\ii(\pn{44}-\pi/2)}+\ee^{\ii(\pn{24}+\pi/2)}+\ee^{\ii(\pn{42}+\pi/2)}\Big],
  \label{eq:read_a}
\end{equation}
where the consistency $|z|\leq1$ equals one for a member of the family. A block with $|z|$ well below one lies outside the family, for example in a different arrangement of its output ports, and $\hat a$ is then not defined.

For the three targets $a=0,\pi/3,2\pi/3$ (0, 1.05 and 2.09\,rad), the phase-aware programs give $\hat a=-0.12$, $1.03$ and $2.02$\,rad on the chip. The intensity-only programs reach the same intensity fidelity but land on other members of the family, $\hat a=4.35$, $1.97$ and $4.66$\,rad, close to the members predicted by the twin before the measurement, $4.37$, $1.99$ and $4.72$\,rad. The target $a=\pi/3$ is the intensity-only program that comes closest to its target ($\Fphi=0.953$), as the twin predicted.

\section{Angle encoding}
\label{sec:encoding}

\emph{Input vectors.} With the MZI convention of Eq.~\eqref{eq:mzi_matrix}, light entering port 2 is split by the chain A1 (ports 1,2), B1 (ports 2,3) and C2 (ports 3,4), giving the real amplitudes of Table~\ref{tab:encoding} up to the phases of the paths, which are set by the phase shifters of D1, E1 and E2. For the phase encoding, B1 splits the light into ports 2 and 3, C1 and C2 split it again into ports 1--4 with equal powers, and the phase shifters of D1, E1 and E2, held in the bar state, write the phases of ports 2, 1 and 3 relative to port 4. With $\theta_k=k\pi x$, the amplitude encoding covers a path through the four output distributions that starts and ends at single-port inputs ($x=0$ and $1$), where the output does not depend on the phase of the block.

\emph{Encoder solution.} The encoder is solved through the twin for each input and program, with the block in the bar state, as described in Methods. The target field at the block, $\bm N^{-1}\bm D_{\rm in}^{\ast}\bm v$, compensates the input port phases and the column transmissions of the programmed block, so that the intended input $\bm v$ reaches the block's transformation $\U$. The calibration uses only the twin and is fixed before the measurement; it does not use measured data.

\emph{Encoder and drift checks.} With the block in the bar state, the outputs read the powers $|v_k|^2$ prepared by the encoder (the phases are not visible in this configuration). Across all angle-encoding measurements, the measured encoder powers agree with the intended ones with a mean fidelity of $0.9975$ (minimum $0.988$, 286 checks) for amplitude encoding and $0.9979$ (minimum $0.993$, 156 checks) for phase encoding. A fixed reference configuration, repeated 76 times, agrees with the twin with a mean fidelity of $0.9982$.

\emph{Results by target set.} Supplementary Table~\ref{tab:encoding_results} lists the mean output fidelities of Fig.~\ref{fig:encoding}(c) by target set and encoding, averaged over all inputs.

\begin{table}[htbp]
\centering
\caption{Mean output fidelity of the angle-encoding measurements, averaged over the inputs (51 values of $x$ for amplitude encoding and 30 random draws for phase encoding per program). Haar: ten Haar-random targets; Hadamard: $F_4(a)$, $a=0,\pi/3,2\pi/3$.}
\label{tab:encoding_results}
\small\setlength{\tabcolsep}{4pt}
\begin{tabular}{llccccc}
\hline
Targets & Encoding & $\lambda_{\rm ph}$ & Inputs & Measured vs target & Twin vs target & Measured vs twin\\
\hline
Haar & amplitude & 0   & 510 & 0.9472 & 0.9482 & 0.9966\\
Haar & amplitude & 0.5 & 510 & 0.9965 & 0.9998 & 0.9966\\
Haar & phase     & 0   & 300 & 0.9355 & 0.9398 & 0.9955\\
Haar & phase     & 0.5 & 300 & 0.9955 & 0.9999 & 0.9956\\
Hadamard & amplitude & 0   & 153 & 0.9402 & 0.9473 & 0.9938\\
Hadamard & amplitude & 0.5 & 153 & 0.9963 & 0.9998 & 0.9965\\
Hadamard & phase     & 0   & 90  & 0.8673 & 0.8750 & 0.9941\\
Hadamard & phase     & 0.5 & 90  & 0.9935 & 0.9999 & 0.9937\\
\hline
\end{tabular}
\end{table}

\section{Direction-of-arrival estimation}
\label{sec:doa}

\emph{Array and beams.} For a plane wave arriving at angle $\vartheta$ from broadside on a linear array of four elements with spacing $d=\lambda/2$, element $k$ receives the phase $(k-1)\,2\pi d\sin\vartheta/\lambda=(k-1)\Delta$, $\Delta=\pi\sin\vartheta$, so the normalised array signal is $v_k(\vartheta)=\ee^{\ii(k-1)\Delta}/2$. A Butler matrix~\cite{ButlerLowe1961} applies the DFT to these signals. With $F_4(\pi)_{jk}=\tfrac12\ee^{-\ii\pi(j-1)(k-1)/2}$, output $j$ receives
\begin{equation}
  Q_j(\vartheta)=\Big|\frac14\sum_{k=1}^{4}\ee^{\ii(k-1)[\Delta-(j-1)\pi/2]}\Big|^2,
  \label{eq:beams}
\end{equation}
which equals one when $\Delta=(j-1)\pi/2$, i.e.\ $\sin\vartheta=0,\tfrac12,1,-\tfrac12$ for $j=1,\ldots,4$. The beams are therefore equally spaced in $\sin\vartheta$, at $\vartheta=0^\circ$, $30^\circ$, $\pm90^\circ$ and $-30^\circ$. At $\vartheta=\pm90^\circ$, $\Delta=\pm\pi$ give the same signals, so the two sides cannot be distinguished there.

\emph{Phase sensitivity.} A block $F_4(\pi+\delta)$ of the family of Eq.~\eqref{eq:f4} has the same powers as the DFT but shifts the beams of outputs 2 and 4. Decoding its outputs with the ideal DFT beams then gives an angle error that depends only on $\delta$. The intensity-only programs land on members with large $\delta$ (Supplementary Table~\ref{tab:doa}) or, for some starts, outside the family, and their direction-of-arrival (DOA) errors follow the twin's prediction.

\emph{Programs.} The DFT is programmed from five random starts with $\lambda_{\rm ph}=0$ and $0.5$. Supplementary Table~\ref{tab:doa} lists, for each program, the deviation $\delta=\hat a-\pi$ read from the bridge-chain measurement of the block with Eq.~\eqref{eq:read_a}, the aligned phase fidelity against the DFT (Methods), and the measured and twin-predicted DOA errors. As controls, three targets $F_4(\pi+\delta)$ with $\delta=30^\circ$, $60^\circ$ and $90^\circ$ were programmed with $\lambda_{\rm ph}=0.5$; the measured blocks give $\delta=31.8^\circ$, $60.5^\circ$ and $90.5^\circ$, and their DOA errors, $3.9^\circ$, $23.6^\circ$ and $34.6^\circ$, are those expected for blocks that are deliberately not the DFT.

\emph{Encoder check.} With the block in the bar state (BAR), the encoder powers agree with the intended ones with a mean fidelity of $0.9981$ (minimum $0.994$, 160 checks).

\begin{table}[htbp]
\centering
\caption{DFT programs of the DOA experiment. $\delta$: deviation of the measured block from the DFT, read with Eq.~\eqref{eq:read_a}; ``--'' marks blocks with consistency $|z|<0.9$, which lie outside the family $F_4(a)$. $\Fphi$: aligned phase fidelity of the measured block against the DFT. DOA errors are mean absolute errors for $|\vartheta|\leq60^\circ$.}
\label{tab:doa}
\small
\begin{tabular}{cccccc}
\hline
Start & $\lambda_{\rm ph}$ & $\delta$ (deg) & $\Fphi$ vs DFT & Measured error (deg) & Twin error (deg)\\
\hline
1 & 0   & $-92.8$  & 0.851 & 57.4 & 59.9\\
2 & 0   & $-108.8$ & 0.784 & 22.1 & 21.0\\
3 & 0   & --       & 0.446 & 29.0 & 26.5\\
4 & 0   & $-70.1$  & 0.884 & 17.8 & 15.2\\
5 & 0   & --       & 0.426 & 19.3 & 19.8\\
1 & 0.5 & $2.5$    & 0.9988 & 1.8 & 0.2\\
2 & 0.5 & $1.3$    & 0.9993 & 2.2 & 0.2\\
3 & 0.5 & $-2.2$   & 0.9995 & 2.1 & 0.2\\
4 & 0.5 & $-4.4$   & 0.9990 & 2.2 & 0.2\\
5 & 0.5 & $-7.3$   & 0.9986 & 2.0 & 0.2\\
\hline
\end{tabular}
\end{table}

\section{Coherent inputs into \texorpdfstring{$5\times5$}{5x5} and \texorpdfstring{$6\times6$}{6x6} blocks}
\label{sec:coherent_large}

\begin{figure}[htbp]
\centering
\includegraphics[width=\textwidth]{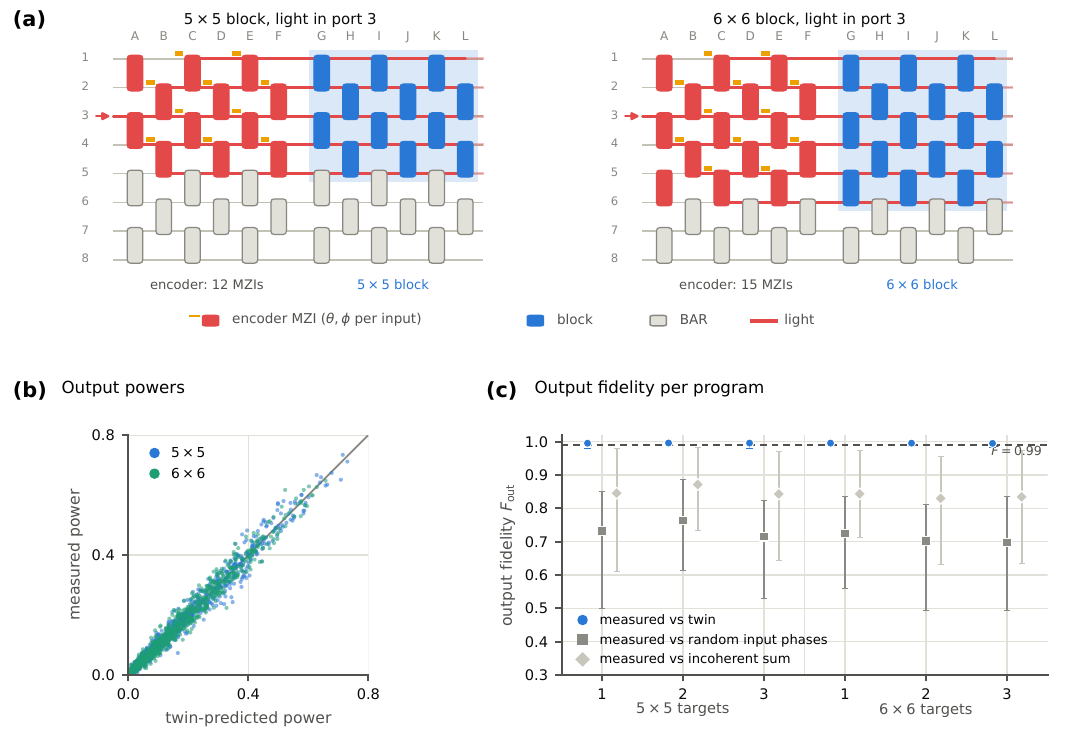}
\caption{Coherent inputs into $5\times5$ and $6\times6$ blocks. (a) Encoder layouts: the block (blue) occupies layers G--L on ports 1--$m$; all MZIs of layers A--F within these ports (red) form the encoder, whose internal and external phases are solved for each input; light enters port 3. The external phases of layer A are not driven. (b) Measured against twin-predicted output power for every output of the 50 random inputs of each of the six programs. (c) Output fidelity of the measurement against the twin and against two predictions that use only the powers of the target: random input phases and an incoherent sum. Markers, mean over the 50 inputs; bars, range. The dashed line marks $F_{\rm out}=0.99$.}
\label{fig:coherent_large}
\end{figure}

To test the coherent-input prediction for larger blocks, three Haar-random targets are programmed for each size, as $5\times5$ and $6\times6$ blocks on ports 1--5 and 1--6 (layers G--L), from a Clements warm start with $\lambda_{\rm ph}=0.5$. Light enters port 3, and all MZIs in layers A--F within the ports of the block serve as the encoder, solved as in Methods: 12 MZIs for the $5\times5$ block and 15 for the $6\times6$ block [Supplementary Fig.~\ref{fig:coherent_large}(a)]. The same construction applies to other block sizes as long as enough layers remain in front of the block for the light from one port to reach all of its inputs with set amplitudes and phases; for the $4\times4$ blocks of Fig.~\ref{fig:encoding}, a subset of these MZIs sets either the amplitudes or the phases. For each size, 50 random input vectors are drawn from a complex Gaussian distribution and normalised, and the same vectors are used for the three targets. The measured outputs are compared with the twin's prediction and with two predictions that use only the powers of the target: one with random input phases, $|T\,\mathrm{diag}(\ee^{\ii\rho})\bm v|^2$ averaged over random $\rho$, and an incoherent sum, $|T|^2|\bm v|^2$.

The measured outputs agree with the twin with a mean fidelity of $0.9957$ for $5\times5$ and $0.9954$ for $6\times6$ (minimum $0.979$), and with the target with $0.9952$ and $0.9947$ [Supplementary Fig.~\ref{fig:coherent_large}(b,c) and Supplementary Table~\ref{tab:coherent}].

 The power-only models reach only $0.70$--$0.76$ (random phases) and $0.83$--$0.87$ (incoherent sum), so the outputs depend strongly on the phase of the block. The encoder alone, measured with the block at BAR, reaches $0.9953$ ($5\times5$) and $0.9966$ ($6\times6$). The same blocks measured with the bridge chain give phase fidelities against the target of $0.989$ ($5\times5$) and $0.994$ ($6\times6$).

\begin{table}[htbp]
\centering
\caption{Coherent inputs into $5\times5$ and $6\times6$ blocks; 50 random inputs per program. Mean output fidelity of the measurement against the twin and the target, of the twin against the target, and of the measurement against two power-only models.}
\label{tab:coherent}
\small\setlength{\tabcolsep}{4pt}
\begin{tabular}{lccccc}
\hline
Program & Meas.\ vs twin & Meas.\ vs target & Twin vs target & Random phases & Incoherent\\
\hline
$5\times5$, target 1 & 0.9954 & 0.9950 & 0.9998 & 0.732 & 0.846\\
$5\times5$, target 2 & 0.9960 & 0.9954 & 0.9998 & 0.763 & 0.872\\
$5\times5$, target 3 & 0.9957 & 0.9951 & 0.9998 & 0.715 & 0.843\\
$6\times6$, target 1 & 0.9956 & 0.9953 & 0.9999 & 0.724 & 0.843\\
$6\times6$, target 2 & 0.9955 & 0.9949 & 0.9997 & 0.702 & 0.830\\
$6\times6$, target 3 & 0.9949 & 0.9940 & 0.9996 & 0.698 & 0.834\\
\hline
\end{tabular}
\end{table}

\section{Noise floor and thermal crosstalk}
\label{sec:noise}

The measured outputs differ from the twin's prediction by about $4\times10^{-3}$ in output infidelity (Figs.~\ref{fig:encoding} and \ref{fig:doa}). To determine whether this residual is set by the measurement or by the model, we measured the repeatability of the readings and the effect of heaters that carry no light, using configurations of the $5\times5$ and $6\times6$ coherent scan.

\emph{Noise floor.} Twenty configurations are each measured ten times in round-robin order, so that every repeat sets all currents afresh. The repeat infidelity of a reading, $1-F$ against the mean of the other readings of the same configuration, has a mean of $5.6\times10^{-5}$ (median $2.6\times10^{-5}$). For the same readings, the infidelity against the twin has a mean of $4.5\times10^{-3}$ (median $3.1\times10^{-3}$), 81 times larger. The total detected power varies by $0.6\,\%$ between repeats, which the normalisation removes. The measured--twin residual is therefore set by the model, not by the measurement noise.

\emph{Thermal crosstalk.} The twin has no thermal crosstalk term: each heater sets only the phase of its own MZI. For each of six base configurations, we identified with the twin the active heaters whose change alone does not alter the predicted outputs, grouped them by their distance from the light path (near: one mode; mid: two to three modes; far: four or more modes; all), and switched each group to a high-dissipation state, measuring the base configuration before and after each group. Supplementary Table~\ref{tab:xtalk} lists the change of the outputs, $1-F$ against the mean of the two neighbouring base readings. The changes are at most $2\times10^{-4}$, about 4\,\% of the measured--twin residual; they reach the base-to-base scatter for the near and far groups, exceed it by a factor of three to six for the mid and all groups, and do not grow towards the light path. Thermal crosstalk not included in the twin therefore does not explain the residual. In an earlier test, adding pairwise and global crosstalk terms to the twin did not improve its test fidelity, and the fitted coefficients were below 0.5\,\% for 90\,\% of the pairs.

\begin{table}[htbp]
\centering
\caption{Thermal crosstalk test. Heaters that carry no light are switched to a high-dissipation state in groups by their distance from the light path; six base configurations per group. The twin predicts no change by construction.}
\label{tab:xtalk}
\begin{tabular}{lcccc}
\hline
Group & Heaters & $\sum|\Delta I^2|$ (mA$^2$) & Change $1-F$ & Base-to-base $1-F$\\
\hline
near (1 mode)      & 11 & 13.8 & $5.9\times10^{-5}$ & $4.6\times10^{-5}$\\
mid (2--3 modes)   & 23 & 24.9 & $1.5\times10^{-4}$ & $5.2\times10^{-5}$\\
far ($\geq4$ modes) & 23 & 26.7 & $6.3\times10^{-5}$ & $4.9\times10^{-5}$\\
all                & 57 & 65.4 & $2.0\times10^{-4}$ & $3.2\times10^{-5}$\\
\hline
\end{tabular}
\end{table}

\section{Device and experimental setup}
\label{sec:chipnote}

\begin{figure}[htbp]
\centering
\includegraphics[width=\textwidth]{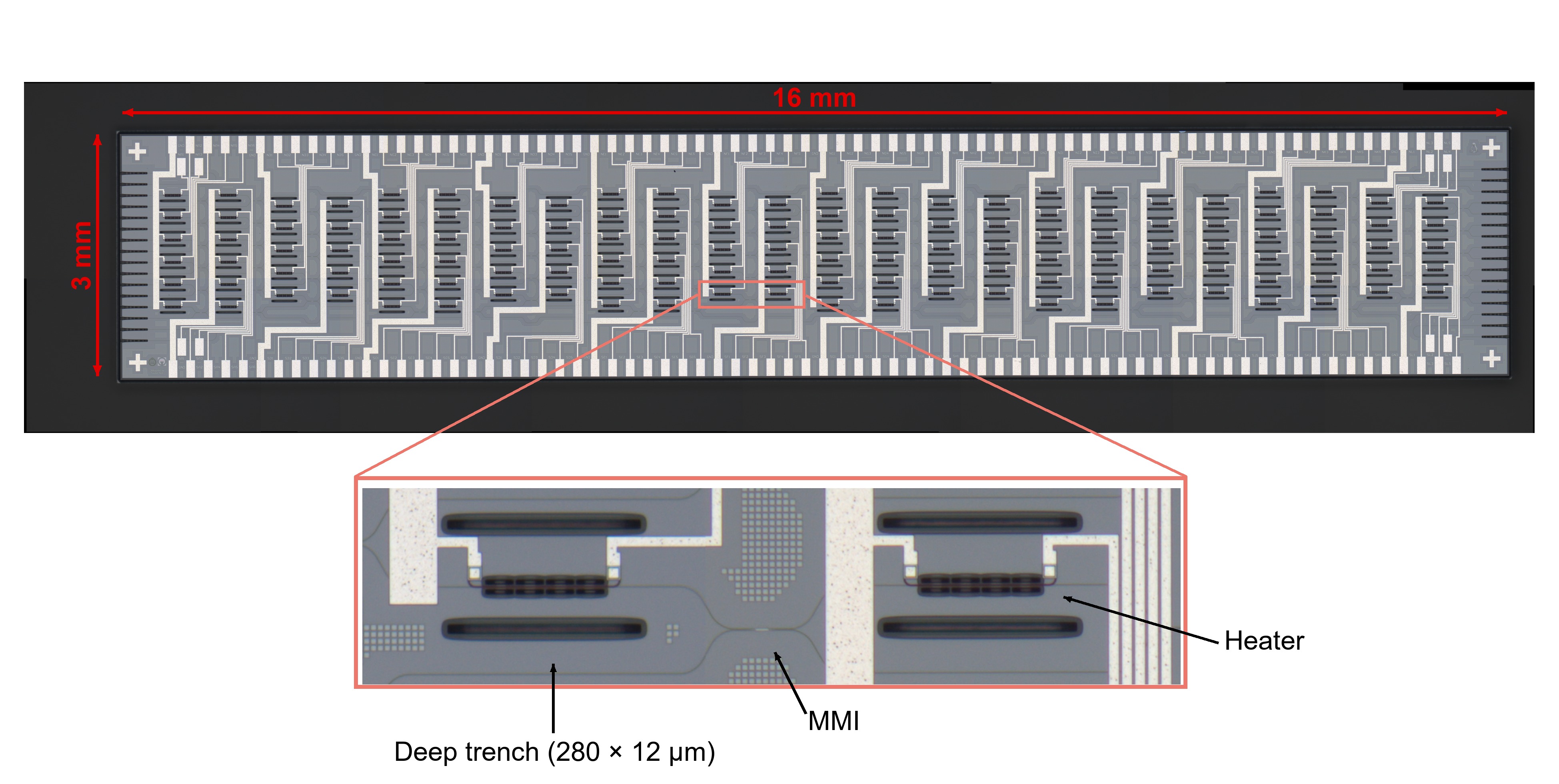}
\caption{Micrograph of the $12\times12$ programmable mesh. Top: full chip, about $16\times3$\,mm$^2$, with the heater contact pads along both long edges. Bottom: enlarged region (red box) showing two MZIs, with an MMI coupler, a heater on one arm and the deep trenches ($280\times12$\,\textmu m$^2$) beside the heated arms.}
\label{fig:chip}
\end{figure}

The processor is a $12\times12$ Clements mesh of 66 MZIs on a silicon-on-insulator chip of about $16\times3$\,mm$^2$ (Supplementary Fig.~\ref{fig:chip}).
Each MZI consists of two MMI couplers and two thermo-optic phase shifters, resistive heaters above the silicon waveguides, one between the couplers ($\theta$) and one on an input arm ($\phi$). Deep trenches of $280\times12$\,\textmu m$^2$ run beside the heated arms to reduce the heat flow to neighbouring waveguides. Light is coupled to and from the chip through polarisation-maintaining V-groove fibre arrays. Light at 1550\,nm from a tunable laser (TOPTICA DLC pro) passes a three-paddle fibre polarisation controller, which launches the transverse-electric mode, and a $1\times12$ microelectromechanical (MEMS) optical switch that selects the input port. A second $1\times12$ MEMS switch routes one output port at a time to a power meter (Thorlabs PM100D), so that the 12 outputs of a setting are read in sequence. The heaters are driven by current drivers (Qontrol Q8iv) with a resolution of 370\,nA, which set all heaters of a configuration in parallel. The chip sits on a thermoelectric cooler, held at 25\,$^\circ$C by a temperature controller (Thorlabs TED200C) with a 10\,k$\Omega$ thermistor. The same setup was used in Refs.~\cite{Cataldo2026PRL,Cataldo2026Floquet}.

\section{Comparison with related work}
\label{app:related}

Supplementary Table~\ref{tab:related} compares this work with related approaches to the model-based characterisation and programming of programmable interferometers. A physical model fitted to single-input output powers yields a complex transfer matrix and has been used to program a 12-mode Clements mesh, with the implemented transformation evaluated through its amplitude fidelity~\cite{Fyrillas2024}. Algebraic and feature-based reconstructions recover the transfer matrix including its phases, but they require intensity-based tomography with controlled input phases or phase-resolved tomography, and have been demonstrated in simulation~\cite{Bantysh2024,Kuzmin2025}. The closest experiment compared the transfer matrices predicted by a model trained on single-input power sweeps with matrices measured using two coherent beams from a dedicated on-chip preparation interferometer, for a four-mode interferometer with three phase shifters that is not universal~\cite{Kondratyev2026}; its comparison after setting the first row and column real is equivalent to the normalised phase used here. On the same processor, the relative phases of a three-mode output state were read with an MZI of the mesh behind the block, set to 50:50, by sweeping one external phase; the phase reference of this sweep comes from a calibration of each phase shifter by node isolation and virtual MZIs, so offsets and path phases between the block and the measuring MZI enter the reading unless calibrated~\cite{Xu2026Sync}. Black-box neural-network calibration of $3\times3$ and $4\times4$ meshes has predicted $|U|^2$, and phase has been predicted for a $2\times2$ gate with coherent detection~\cite{RausellCampo2026}. Here, the twin is trained on single-input powers at random settings of all heaters of a universal $12\times12$ mesh, the normalised phase is measured for blocks of different sizes and positions using interference between beams prepared by the MZIs of the mesh itself, and this phase enters the inverse design through the phase term.

\begin{table}[htbp]
\centering
\caption{Comparison with related work on model-based characterisation and programming of programmable interferometers. ``Phase checked on chip'' refers to an experimental comparison of the model-predicted phase with an independent on-chip phase measurement. Exp.: experiment; sim.: simulation.}
\label{tab:related}
\footnotesize
\setlength{\tabcolsep}{3pt}
\renewcommand{\arraystretch}{1.3}
\begin{tabular}{p{1.2cm}p{2.6cm}p{2.5cm}p{3.0cm}p{2.6cm}p{2.5cm}}
\hline
{\raggedright Ref.\par} & {\raggedright Device\par} & {\raggedright Training data\par} & {\raggedright Phase checked on chip\par} & {\raggedright Validation scale\par} & {\raggedright Phase used in programming\par} \\
\hline
{\raggedright Ref.~\cite{Fyrillas2024}\par} & {\raggedright $12\times12$ Clements mesh, 66 MZIs, 126 phase shifters (exp.)\par} & {\raggedright Single-input powers\par} & {\raggedright No; amplitude fidelity\par} & {\raggedright Full $12\times12$, 100 Haar targets\par} & {\raggedright No; corrected Clements decomposition\par} \\[5pt]
{\raggedright Ref.~\cite{Bantysh2024}\par} & {\raggedright Multiport architecture, $N=10$ (sim.)\par} & {\raggedright Intensity-based tomography with controlled input phases\par} & {\raggedright Not applicable (sim.)\par} & {\raggedright --\par} & {\raggedright --\par} \\[5pt]
{\raggedright Ref.~\cite{Kuzmin2025}\par} & {\raggedright Multiport architecture, $N\le5$ (sim.)\par} & {\raggedright Phase-resolved tomography\par} & {\raggedright Not applicable (sim.)\par} & {\raggedright --\par} & {\raggedright Full-matrix objective (sim.)\par} \\[5pt]
{\raggedright Ref.~\cite{Kondratyev2026}\par} & {\raggedright 4-mode multiport interferometer with 3 phase shifters, not universal (exp.)\par} & {\raggedright Single-input powers, one heater swept at a time\par} & {\raggedright Yes; two beams from a dedicated on-chip preparation interferometer, average matrix fidelity 99.6\,\%\par} & {\raggedright Full $4\times4$, 100 random settings (not Haar)\par} & {\raggedright No; output power distribution\par} \\[5pt]
{\raggedright Ref.~\cite{Xu2026Sync}\par} & {\raggedright $12\times12$ Clements mesh; $3\times3$ non-unitary block in a $5\times5$ dilation (exp.)\par} & {\raggedright Per-phase-shifter calibration, validated by output intensities of identity and DFT settings\par} & {\raggedright Output-state phases against the target state; 3\,dB MZI behind the block, phase reference from calibration\par} & {\raggedright One 3-mode output state, two phase differences per step\par} & {\raggedright No; calibrated decomposition of the dilation\par} \\[5pt]
{\raggedright Ref.~\cite{RausellCampo2026}\par} & {\raggedright $3\times3$ and $4\times4$ meshes; $2\times2$ gate (exp.)\par} & {\raggedright $|U|^2$; coherent detection for the $2\times2$ gate\par} & {\raggedright $2\times2$ gate only\par} & {\raggedright $3\times3$, $4\times4$\par} & {\raggedright No; inverse network on $|U|^2$\par} \\[5pt]
{\raggedright This work\par} & {\raggedright $12\times12$ Clements mesh, 66 MZIs, 126 driven phase shifters (exp.)\par} & {\raggedright Single-input powers, all heaters random\par} & {\raggedright Yes; two beams from MZIs of the mesh itself, normalised phase $\varphi^{\mathrm{norm}}$\par} & {\raggedright $m\times m$ blocks, $m=3$--$10$, 46 configurations; 13 paired $4\times4$ targets; coherent inputs and DOA\par} & {\raggedright Yes; phase term $\lambda_{\rm ph}$, paired comparison\par} \\
\hline
\end{tabular}
\end{table}

\FloatBarrier

\bigskip\noindent\emph{References cited in the Supplementary Information refer to the reference list of the main text.}

\end{document}